\documentclass[sn-basic, iicol]{sn-jnl}

\usepackage{graphicx}%
\usepackage{natbib}
\setcitestyle{comma, numbers,sort&compress, super}
\usepackage{array}
\usepackage{multirow}%
\usepackage{subfig}
\usepackage{caption}
\usepackage{tabularx}
\usepackage{amsmath,amssymb,amsfonts}%
\usepackage{amsthm}%
\usepackage{mathrsfs}%
\usepackage[title]{appendix}%
\usepackage{xcolor}%
\usepackage{textcomp}%
\usepackage{manyfoot}%
\usepackage{booktabs}%
\usepackage{algorithm}%
\usepackage{algorithmicx}%
\usepackage{algpseudocode}%
\usepackage{listings}%
\usepackage{arydshln}
\usepackage{pifont}

\theoremstyle{thmstyleone}%
\theoremstyle{thmstyletwo}%

\theoremstyle{thmstylethree}%

\newcolumntype{P}[1]{>{\centering\arraybackslash}p{#1}}

\newcommand{\eg}{\textit{e.g}.}

\newcommand{\etc}{\textit{etc}}

\newcommand{\tabincell}[2]{\begin{tabular}{@{}#1@{}}#2\end{tabular}}

\begin{document}

\title[Beyond Image Prior: Embedding Noise Prior into Latent Space of Conditional Denoising Transformer]{Beyond Image Prior: Embedding Noise Prior into Latent Space of Conditional Denoising Transformer}


\author[1]{\fnm{Yuanfei} \sur{Huang}}\email{yfhuang@bnu.edu.cn}

\author*[1]{\fnm{Hua} \sur{Huang}}\email{huahuang@bnu.edu.cn}

\affil[1]{\orgdiv{School of Artificial Intelligence}, \orgname{Beijing Normal University}, \orgaddress{\city{Beijing}, \postcode{100875}, \country{China}}}


\abstract{Existing learning-based denoising methods typically train models to generalize the image prior from large-scale datasets, suffering from the variability in noise distributions encountered in real-world scenarios. In this work, we propose a new perspective on the denoising challenge by highlighting the distinct separation between noise and image priors. This insight forms the basis for our development of conditional optimization framework, designed to overcome the constraints of traditional denoising framework. To this end, we introduce a Locally Noise Prior Estimation (LoNPE) algorithm, which accurately estimates the noise prior directly from a single raw noisy image. This estimation acts as an explicit prior representation of the camera sensor's imaging environment, distinct from the image prior of scenes. Additionally, we design an auxiliary learnable LoNPE network tailored for practical application to sRGB noisy images. Leveraging the estimated noise prior, we present a novel Conditional Denoising Transformer (Condformer), by incorporating the noise prior into a conditional self-attention mechanism. This integration allows the Condformer to segment the optimization process into multiple explicit subspaces, significantly enhancing the model's generalization and flexibility. Extensive experimental evaluations on both synthetic and real-world datasets, demonstrate that the proposed method achieves superior performance over current state-of-the-art methods. The source code is available at \url{https://github.com/YuanfeiHuang/Condformer}.}




\maketitle

\section{Introduction}\label{sec1}
Image denoising, a fundamental aspect of low-level vision, is garnering increased interest due to its significant applications in computational photography and computer vision. The primary goal of image denoising is to mitigate the impact of unwanted noise in noisy observations, thereby enhancing image quality for either aesthetic enhancement or to facilitate subsequent processing tasks.

\begin{figure*}[!t]
	\centering
	\captionsetup[subfloat]{justification=centering}
	\subfloat[]{
		\centering
		\includegraphics[height=0.18\textheight]{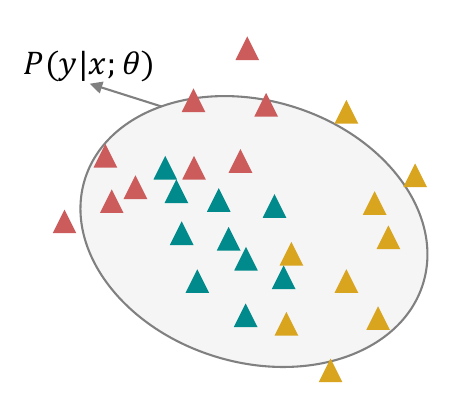}
		\label{fig:optim_a}
	}\hspace{1cm}
	\subfloat[]{
		\centering
		\includegraphics[height=0.18\textheight]{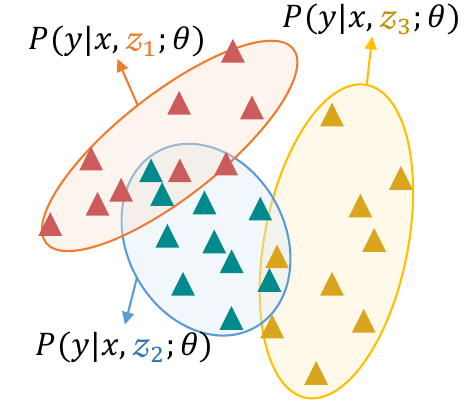}
		\label{fig:optim_b}
	}
	\subfloat{
		\centering
		\includegraphics[height=0.18\textheight]{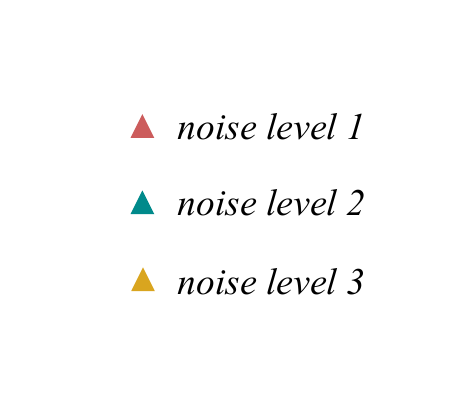}
	}
	\caption{Illustration of denoising model optimization: (a) unconditional optimization space with image prior $x$ \citep{ZamirS2022CVPR,GuoH2024ECCV}; (b) conditional optimization with noise prior $z$ and image prior $x$ in this work.}
	\label{fig:optimization}
\end{figure*}
In general, natural images embody strong priors for visual perception \citep{UlyanovD2018CVPR,LehtinenJ2018ICML}, such as repetitive textures and continuous edges, which are more readily inferred than the seemingly random presence of noise. Thus, a prevailing strategy among many existing learning-based denoising methods \citep{ZhangK2017TIP,ZhangK2018TIP,ZamirS2022CVPR,MeiY2023IJCV,GuoH2024ECCV} involves developing a unified model capable of generalizing from a vast collection of noisy-clean pairs. This process typically formulates optimization as:
\begin{equation}
	\hat{\theta} = \arg\max_{\theta}\mathbb{E}_x\log P(y|x;\theta)
\end{equation}
where $\theta$ represents the model parameters, $x$ and $y$ denote the noisy observation and clean target, respectively. In this context, the noise map is treated as an additive mask over the clean image, with the ultimate aim of deducing the underlying image prior. However, the noise prior, which is crucial for distinguishing between various noise distributions, is often overlooked.

This framework faces limitations in real-world scenarios due to two primary challenges:

1) The difficulty and cost associated with gathering large-scale noisy-clean image datasets. Learning sophisticated image priors necessitates a model with substantial capacity, which is hindered by the challenges of acquiring clean images. Clean image acquisition typically requires long exposures in static scenes \citep{AbdelhamedA2018CVPR,AnayaJ2017JVCIR,PlotzT2017CVPR} or involves complex alignment procedures \citep{AbdelhamedA2018CVPR}.

2) The inefficiency and incompleteness of an unconditional optimization space. Conventional learning-based denoising methods, which focus solely on image priors, are inherently unconditional. However, as illustrated in Fig.~\ref{fig:optim_a}, these methods attempt to learn and generalize the image prior from numerous noisy-clean samples, resulting in an optimization space that is both overly broad, encompassing unnecessary scenarios, and simultaneously incomplete, missing critical outlier cases.

To address these issues, we propose segmenting the singular unconditional optimization space into multiple subspaces by incorporating reliable noise priors alongside the complicated image priors. As illustrated in Fig.~\ref{fig:optim_b}, this approach recognizes that a noisy observation is influenced both by the scene (image prior) and the imaging environment (noise prior), making it logical to infer the noise prior for optimizing the denoising process. Consequently, the optimization space for a conditional denoising model should comprise various independent and complete subspaces, each conditioned on specific priors, and can be represented as:
\begin{equation}
	\hat{\theta} = \arg\max_{\theta}\sum_{i=1}^{n}\mathbb{E}_{x_i}\log P(y|x_i,z_i;\theta)
	\label{eq:cond_optim}
\end{equation}
where $\{z_i\}^n_{i=1}$ represents the noise prior.

Distinct from traditional models that rely on given conditional embedding (\eg, FFDNet \citep{ZhangK2018TIP}) or implicit noise prediction (\eg, VDN \citep{YueZ2019NeurIPS} and CVF-SID \citep{NeshatavarR2022CVPR}), a conditional denoising model must adaptively estimate an explicit noise prior from a single noisy observation and distinctly address the noise and image priors based on their independence, rather than concatenate the image and noise parameter directly as their mismatching is a critical obstacle for improving denoising performance. 
In essence, as depicted in Fig.~\ref{fig:optim_b}, the principle of a conditional denoising model lies in its ability to navigate the generation of pixels by harnessing implicit natural image priors to shape the optimization landscape, while also leveraging explicit sensor noise priors to precisely target the optimization's focus. 

Building upon this concept, we introduce a novel approach for explicit noise prior estimation from a single noisy observation, termed \textbf{Lo}cally \textbf{N}oise \textbf{P}rior \textbf{E}stimation ({\bf LoNPE}), and develop a \textbf{Con}ditional \textbf{d}enoising Trans\textbf{former} ({\bf Condformer}) that incorporates this noise prior. This integration allows for the segmentation of the entire optimization space into distinct, explicit optimization subspaces. 
Our main contributions are summarized as follows:
\begin{itemize}
	\item By rethinking the imaging mechanism in physics, we offer a new perspective on image denoising, highlighting the independence between noise and image priors. This distinction is crucial for conditional optimization, particularly within the context of real-world scenarios.
	
	\item We introduce an innovative LoNPE algorithm for estimating noise prior from raw noisy image. This method effectively captures the characteristics of sensor noise, providing an explicit prior for conditional optimization. Additionally, we present a learnable LoNPE network, tailored for practical application with only single sRGB noisy observation.
	
	\item By exploring the noise statistics concealing in the latent space, we propose a novel Condformer that leverages the estimated noise prior within a conditional self-attention module. This design represents a pioneering effort to incorporate prior knowledge into a Transformer-like architecture for denoising, and alleviates the mismatching issue in existing conditional denoisers. 
	
	\item Quantitative and qualitative experiments demonstrate the superior performance of our LoNPE algorithm and Condformer model across various real and synthetic noise analysis and image denoising tasks.
	
\end{itemize}

The rest of this paper is organized as follows: Section~\ref{sec:related work} reviews related work. Section~\ref{sec:methods} presents our noise prior estimation and conditional image denoising methods. Qualitative and quantitative experiments are reported and analyzed in Section~\ref{sec:experiments}. Finally, Section~\ref{sec:conclusion} concludes our work and discusses the limitation and future work.

\section{Related Work}~\label{sec:related work}
As the core goal of this paper is to explore a conditional denoising Transformer with explicit noise prior, we next mainly introduce the advances in the fields of noise modeling, image denoising, and vision Transformer, respectively.

\subsection{Noise Modeling} 
In general, the signal-independent Gaussian distribution is regarded as a theoretically common hypothesis of noise modeling, and has derived numerous supervised image denoising methods to handle the widely-used additive white Gaussian noise (AWGN). However, the real noise model is more sophisticated.
Particularly, due to the characteristics of imaging sensor, the practical noise could be explicitly modeled on raw sensors, and the corresponding raw sensor noise commonly consists of the signal-dependent shot noise and the signal-independent read noise. 

Typically, the Poisson-Gaussian noise model \citep{FoiA2008TIP} was employed to characterize the distribution of this raw sensor noise and has inspired numerous advances in realistic noise synthesis and real image denoising \citep{LiuX2014TIP,GuoS2019CVPR,WangY2020ECCV}. Beyond the hypothesis on building the distribution of noise, to match the noise model in more complex imaging environment on various devices, multi-frame calibration \citep{WangY2020ECCV} and variance-stabilizing transformations \citep{MakitaloM2011TIP,MakitaloM2014TIP,LiD2022IJCV} technologies were presented to refine the noise model for different sensors. Furthermore, some works attempt to employ the calibrated noise parameter of sensor to explicitly synthesize noisy samples \citep{WangY2020ECCV,WeiK2022TPAMI,FengH2024TPAMI}, or guide the network optimization \citep{NeshatavarR2022CVPR,YueZ2024TPAMI} for training a denoising model. Instead of explicitly modeling noise from a certain distribution, learnble noise modeling methods recently have been raised with the development of generative models, such as variational bayes \citep{ZhengD2022TPAMI}, generative adversarial networks \citep{ChangK2020ECCV}, and normalizing flows \citep{MalekyA2022CVPR}.

\subsection{Image Denoising}
After decades of development, image denoising methods are generally divided into model-based and learning-based. 
Model-based denoising methods aim to model the characteristics of natural images as a regularization prior to iteratively optimize a well-designed model. The representative regularization priors include total variation \citep{RudinL1992}, sparsity \citep{WenB2015IJCV}, non-local self-similarity \citep{BuadesA2005CVPR,DabovK2007BM3D}, external statistical priors \citep{XuJ2018TIP}, and Huber function~\citep{SongL2024IJCV}.

On the other hands, learning-based methods attempt to reconstruct a clean image from the noisy observation with an end-to-end learnable model, which is trained from large-scale noisy-clean pairs. 
\subsubsection{Non-blind image denoising}
Initially, with the development of convolutional neural networks (CNN), CNN-based denoising methods \citep{ZhangK2017TIP,ZhangK2018TIP,AnwarS2019ICCV,MeiY2023IJCV,ZhangY2021TPAMI,ZamirS2021CVPR,ZamirS2022TPAMI,PanJ2022IJCV} have received significant advances in learning an end-to-end mapping from the noisy observations to clean targets. In essence, due to the nature of CNN in local visual perception, the key of these CNN-based denoising methods is to learn how a pixel is generated from a corrupted one and its neighbors in local perception region, namely, to learn the image prior in local perceptions, such as details in texture, smoothness in flat.
Nevertheless, some contextual image priors are difficult to capture in local perceptions, \eg, objects, structural information, edges and repeated textures. Recently, considering these image priors in non-local or global visual perception, several Transformer-based \citep{ChenH2021CVPR,LiangJ2021ICCVW}, MLP-based \citep{Tolstikhin2021NeurIPS,TuZ2022CVPR} and Mamba-based \citep{GuoH2024ECCV} denoising methods have attracted increasing attentions and achieved remarkably superior performances against other existing CNN-based methods. In particular, to capture long-range feature dependencies, pyramid \citep{MeiY2023IJCV}, rectangle-window \citep{ZhengC2022NeurIPS}, chaotic-window \citep{XiaoJ2023ICML}, sparse \citep{ZhangJ2023ICLR} and anchored-stripe \citep{LiY2023CVPR} self-attention mechanisms have been explored. Yet capturing spatial correspondence commonly causes quadratically increasing computational loads as the resolution increases, then efficient Transformer-based denoisers recently achieve growing concerns. Specifically, hierarchical U-shape architecture \citep{WangZ2022CVPR} and channel-wise self-attention mechanism \citep{ZamirS2022CVPR} were proposed to reduce the unbearable computational loads from increasing spatial resolutions. 
\subsubsection{Blind image denoising}
Except for the evolution of denoiser architectures, a blind denoising strategy with stronger generalization is essential for practical applications due to the unknowability and the diversity of noise in real scene. In the type of aforementioned supervised learning-based methods, numerous noisy-clean pairs with various noise levels are employed to train an unified denoiser\citep{ZhangK2017TIP,ZhangK2021TPAMI,ZamirS2022CVPR,CuiY2024TPAMI}. To improve the performance, conditional embedding like noise variance map is concatenated with the noisy image in the head of denoiser, to guide model handling a specific given~\citep{ZhangK2018TIP} or predicted \citep{YueZ2024TPAMI} noise level. 
However, mismatch of image and noise level is a critical obstacle of these methods for denoising performance.

Besides, to adapt for real scenes with only noisy observations, self-supervised denoising methods \citep{LehtinenJ2018ICML,NeshatavarR2022CVPR} were raised by learning implicit representation from image priors. However, they often fall short in scenarios requiring high accuracy, complex noise handling, and stable convergence.

\subsection{Vision Transformer}
The self-attention mechanism \citep{VaswaniA2017NeurIPS} in Transformers facilitates learning long-range dependencies, leading to significant success in computer vision tasks \citep{Dosovitskiy2021ICLR,LiuZ2021ICCV}. ViT \citep{Dosovitskiy2021ICLR} demonstrated the Transformer's effectiveness in non-local visual perception and high-accuracy object recognition, sparking the development of more effective and efficient Transformer architectures for various vision tasks like object detection \citep{LiuZ2021ICCV,HongW2024IJCV}, semantic segmentation \citep{ZhangB2022NeurIPS,ZhangL2024IJCV}, and low-level vision \citep{ChenH2021CVPR,ZamirS2022CVPR,MeiY2023IJCV}.

Especially in low-level vision, self-attention mechanism was firstly utilized to transfer relevant textures in reference-based super-resolution \citep{YangF2020CVPR}. More generally, IPT \citep{ChenH2021CVPR} later introduced a multi-task image processing model using standard Transformer architecture with tokenized inputs.
By integrating the advantage of local attention mechanism of CNN and long-range dependency of Transformer, Swin Transformer \citep{LiuZ2021ICCV} was proposed by introducing the	shifted window scheme and was applied into image restoration tasks \citep{LiangJ2021ICCVW}. To alleviate the limitation of Swin Transformer in receptive fields, cross aggregation Transformer (CAT) \citep{ZhengC2022NeurIPS} was proposed by aggregating features cross different windows to expand the receptive field. In addition, attention retractable Transformer (ART) \citep{ZhangJ2023ICLR} was presented to capture local and global receptive field simultaneously. GRL \citep{LiY2023CVPR} was proposed to explicitly model image hierarchies in global, regional, and local range dependencies.
Despite their effectiveness, these Transformers are computationally intensive as calculating the spacial cross-covariance of large-scale tokens. To address this, Uformer \citep{WangZ2022CVPR} was proposed by building a hierarchical U-shape architecture with locally-enhanced window Transformer blocks. Besides, by calculating the channel-wise cross-covariance, an efficient Restormer \citep{ZamirS2022CVPR} was proposed and achieved state-of-the-art performances in several image restoration tasks.

Nonetheless, these existing Transformers rely on large-scale datasets with perfect labels and unconditional optimization, which is commonly redundant. Instead, inspired by the conditional text generation with Transformer \citep{Hosseini2020NeurIPS,ZhengJ2024IJCV}, we aim to explore a conditional Transformer for image denoising.

\section{Method}~\label{sec:methods}
In an imaging pipeline with a photosensor, the target imaging scene is formulated as incident lights hitting the camera sensor array and then transformed into digital responses for imaging. In this section, we first introduce the noise formation model in an imaging sensor and generalize the independence of noise and image priors, then describe the proposed LoNPE algorithm/network for noise prior estimation and the Condformer architecture for conditional denoising. 

\subsection{Preliminary on Noise Prior}
\subsubsection{Noise Formation Model}
Although the noise in a processed sRGB image is generally complex to explicitly analyze due to the nonlinearity of image signal processing (ISP), the raw noise formation model of a digital sensor in camera is well understood \citep{BrooksT2019CVPR}. In particular, the raw noise in a camera primarily consists of the {\em shot noise} during photon-to-electron conversion and the {\em read noise} during electron-to-digital conversion \citep{WeiK2022TPAMI}. 

\begin{figure*}
	\centering
	\includegraphics[width=1\linewidth]{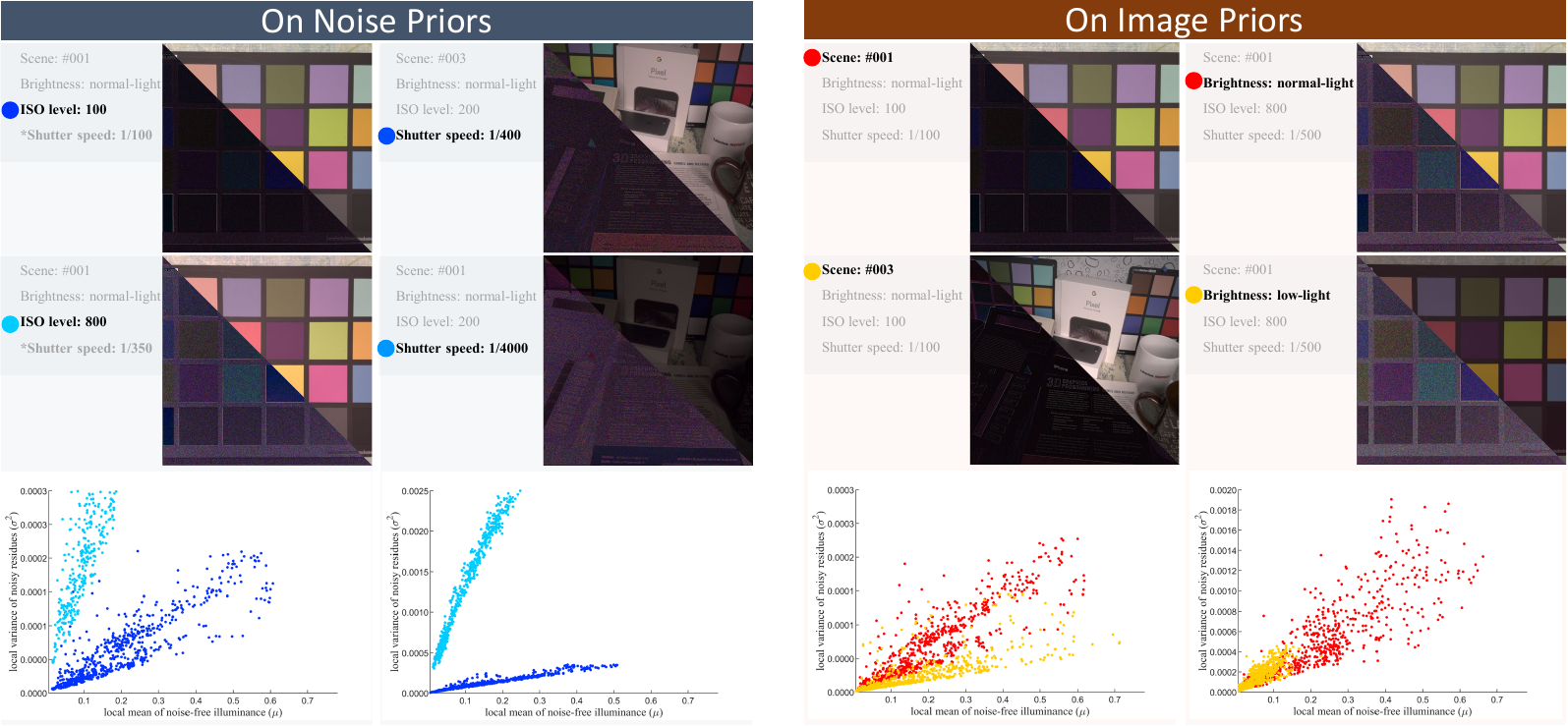}
	\caption{Investigation on independence of the image prior and the noise prior. We select 8 noisy-clean pairs from SIDD training dataset, which are captured with different image prior (related to {\em scene} and {\em brightness}) or noise prior (related to {\em ISO level} and {\em shutter speed}). By investigating the local variance of noisy residues and the local mean of clean image, the statistical results show that the noise prior depends on only the camera settings, but little on the image prior.}
	\label{fig:Noise Prior beyond Image}
\end{figure*}

Specifically, due to the quantum nature of light, the collected noisy photoelectrons can be modeled as Poisson random variable, which follows
\begin{equation}
	(\mathbf{L}+\mathbf{N}_\text{s})\sim \mathcal{P}(\mathbf{L})
\end{equation}
where $\mathbf{L}$ and $\mathbf{N}_\text{s}$ indicate the incident clean photoelectron and the shot noise, respectively. $\mathcal{P}(\cdot)$ denotes the Poisson distribution.

These photoelectrons are subsequently read out as quantizable digital signals, and commonly attached with the read noises $\mathbf{N}_\text{r}$, which can be approximately modeled as Gaussian random variables
\begin{equation}
	\mathbf{N}_\text{r}\sim \mathcal{N}(\mathbf{0}, \sigma^2_\text{r})
\end{equation}
where $\mathcal{N}(\cdot)$ denotes the Gaussian distribution. 

This Poisson-Gaussian can be further treated as a heteroscedastic Gaussian distribution \citep{FoiA2008TIP}, and the final raw digital sensor signals can be formulated as 
\begin{equation}
	\mathbf{I}\sim \mathcal{N}(\mathbf{L}, \sigma^2_\text{s}\cdot \mathbf{L} + \sigma^2_\text{r})
	\label{eq: digital raw signal}
\end{equation}
where $\sigma_\text{s}$ and $\sigma_\text{r}$ indicate the ``{noise prior}'', which depends on the imaging environments, including the camera sensor, and photography settings. 

In this manner, the noise prior plays a significant role in raw sensor noise modeling, and is commonly proportional to the noise level. Particularly, $\mathbf{L}$ indicates the pixel-wise intensity of the target scene illuminance, indicating the ``image prior'', and thereby has nothing with the noise prior. 

\subsubsection{Independence of Noise Prior and Image Prior}\label{sec:independency}
To further illustrate the independence of noise prior and image prior, as shown in Fig.~\ref{fig:Noise Prior beyond Image}, we investigate several raw noisy-clean pairs from SIDD-Medium training dataset, where the noisy observations are captured under different scene illuminances (\eg, scene and brightness) indicating the image prior and different imaging environments (\eg, ISO level and shutter speed) indicating the noise prior. As described above, the noise prior should depend on these imaging environments, and affects the statistical parameter of distribution in pixel-wise variances. 

However, it is infeasible to calculate the pixel-wise variances on a single image. We randomly sample $1000\times16\times16$ local raw noisy-clean patch pairs, and calculate the local variances $\sigma^2$ of the noisy residues and the local means $\mu$ of the clean patches, to further approximate the corresponding pixel-wise statistic of noise. As formulated in Eq.(\ref{eq: digital raw signal}), the pixel-wise variance $\sigma^2$ of noises should be proportional to the intensity $\mu$ of illuminances. Consequently, as the statistical results shown in Fig.~\ref{fig:Noise Prior beyond Image}, its slope and intercept represent $\sigma^2_\text{s}$ and $\sigma^2_\text{r}$, respectively. 

From these observations, on a common sence and environmental brightness, namely with the same image prior, the statistical results of noise show a non-negligible discrepancy for various imaging environments. Specifically, higher ISO level indicates larger sensitivity of the sensor, leading more noises affecting image quality. Faster shutter speed causes lower exposure, and commonly needs to increase the ISO to compensate for the lack of exposure, indirectly increasing the image noise.
Nonetheless, under a common imaging environment (\eg, ISO level or shutter speed) with the same noise prior, different scenes/brightnesses share a statistically similar result of the noise distribution. Thus, this observation indicates the noise prior is beyond and independent on the image prior.

\begin{figure*}
	\centering
	\includegraphics[width=0.9\linewidth]{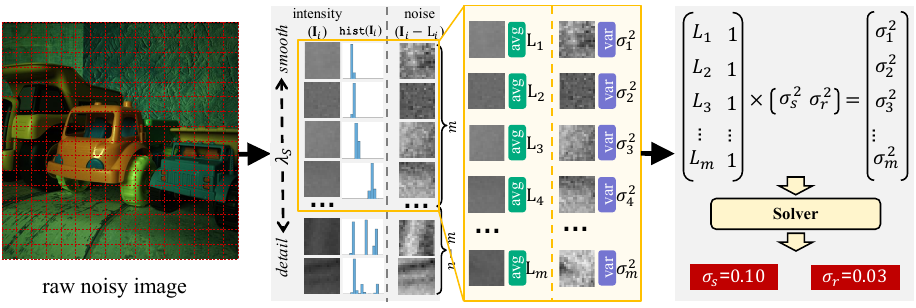}
	\caption{Pipeline of the LoNPE algorithm. The dynamic range of raw noisy image is normalized into $[0,1]$.}
	\label{fig:LoNPE}
\end{figure*}

\subsection{Locally Noise Prior Estimation}\label{sec:LoNPE}

Similar to conventional noise parameter estimation task, which has seen significant progress with methods evolving to address challenges in accuracy and robustness. Different from the Gaussian noise parameter estimation~\citep{ChenG2015ICCV,WangJ2023ICCV,KeR2024TIP,PimpalkhuteV2021TIP}, Poisson-Gaussian noise parameter estimation is more complicated as the noise variance is signal-dependent~\citep{FoiA2008TIP}. Existing methods dedicate to solve the image variance of noisy observation via iterative variance-stabilization~\citep{MakitaloM2014TIP}, maximum likelihood estimation~\cite{LiuX2014TIP}, or learning-based CNN~\citep{ByunJ2021CVPR}. But, challenges remain in handling real cases, especially within strong image prior such as rich textures. 

Specifically in practical usage, as we can only capture a corrupted noisy observation and commonly have little additional information on the environmental illumination, sensor technology, and photography settings, noise prior is basically equivalent to the above noise parameter. To tackle the above challenge, the independence of noise prior and image prior is a critical principle that we can use to estimate the noise prior from a single noisy image. 

As aforementioned, under a common imaging environment, different scenes should have statistically similar noise priors, yet the image prior of a scene generally indicates the features for visual perception, which are commonly represented as texture or edges. The final image commonly has a specific statistical distribution, and each pixel can be formulated as 
\begin{equation}
	\mathbf{I}_i = \Phi(\{\mathbf{I}_j\}_{j\in\mathcal{O}(i)})
\end{equation}
where $\Phi(\cdot)$ indicates image prior model implemented as an onefold denoising model, $\mathcal{O}(i)$ denotes the local neighbors of location $i$. 

Due to the intrinsic sophisticated characteristics of image, the statistical variance of a whole image is less effective to infer neither noise prior nor image prior. 
To effectively separate the noise prior and the image prior, we present a {\bf Lo}cally {\bf N}oise {\bf P}rior {\bf E}stimation ({\bf LoNPE}) algorithm by eliminating the effect of image prior, such as the sophisticated textures, spatially non-local structures and edges, and \etc. 

A primal motivation is to employ the local luminance constancy in a smooth patch, where
\begin{equation}
	\mathbf{I}_i \simeq \mathbb{E}_{j\in\mathcal{O}(i)}(\mathbf{I}_j)
\end{equation}
as shown in Fig. 3, smooth patches exhibit more concentrated histogram distributions, allowing for a more precise characterization of the overall intensity within each patch. This indicates that the image prior is negligible in a local smooth patch and can be effectively represented as the mean value of the pixels within the local region.
Thus, on the local neighbor locations $\mathcal{O}(i)$, the statistical distribution of $\{\mathbf{I}_i\}_{i\in\mathcal{O}(i)}$ should be approximately same, and could be utilized to estimate the noise prior as its independency on the image prior. 

In particular, we firstly preprocess the image to restrict its theoretical value range into $[0, 1]$, then partition it into a group of local patches, and select the smooth patches to eliminate the interference of image prior. Next, we calculate these patches' statistical values of mean and variance, to finally estimate the noise prior ($\sigma_\text{s}$, $\sigma_\text{r}$) with a simple least square optimization solver. 

\subsubsection{LoNPE Algorithm} \label{sec:LoNPE Algorithm}
Due to the variety of sensor bit depth $B$ (\eg, $B=10$ in iPhone 7 camera, $B=14$ in Canon 80D camera), the value range of raw digital sensor image $[0, 2^B]$ might be different. For generally analyzing, we preprocess the raw image by normalizing its theoretical value range into $[0,1]$, as

\begin{equation}
	\mathbf{I} = \mathbf{I} / 2^{B}
\end{equation}

Subsequently, the image is firstly partitioned into $n$ patches $\{\mathbf{I}_i\}^n_{i=1}$ at same size of $\mathcal{O}$. As shown in Fig.~\ref{fig:LoNPE}, on the assumption of local illuminance constancy, we select $m$ smoother patches $\{\mathbf{I}_i\}^m_{i=1}$ as samples for noise prior estimation. In detail, on a smooth patch, the local luminance should be approximated by its statistical mean, as
\begin{equation}
	L_i = \mathbb{E}_{j\in\mathcal{O}(i)}(\mathbf{I}_j)
	\label{eq:LoNPE_mean}
\end{equation}
besides, the statistical variance $\sigma^2$ should contain only noise prior, and is formulated as 
\begin{equation}
	\begin{aligned}
		\sigma^2_i &= L_i\cdot\sigma_\text{s}^2 + \sigma_\text{r}^2\\
		&=\mathbb{E}_{j\in\mathcal{O}(i)}(\mathbf{I}_j-L_i)^2
	\end{aligned}
	\label{eq:LoNPE_var}
\end{equation}

Particularly, to eliminate the image prior, these smooth patches are sampled from the original patch pools $\{\mathbf{I}_i\}^n_{i=1}$ by employing a local smoothness criterion $\lambda_\mathcal{S}$, which is designed to quantify the local smoothness of an heteroscedastic Gaussian distribution, and is formulated as
\begin{equation}
	\lambda_\mathcal{S} = \sigma_i / \sqrt{L_i}
\end{equation}
that, the lower $\lambda_\mathcal{S}$, the higher smoothness in $i$-th local patch.

Based on the independence of noise prior and image prior, all local patches of an image should share a common noise prior ($\sigma_\text{s}$, $\sigma_\text{r}$). Then, we have
\begin{equation}
	\begin{aligned}
		\begin{bmatrix}
			\sigma^2_1\\
			\sigma^2_2\\
			...\\
			\sigma^2_m
		\end{bmatrix}
		&=\begin{bmatrix}
			L_1\\
			L_2\\
			...\\
			L_m
		\end{bmatrix} \cdot \sigma^2_\text{s} + \sigma^2_\text{r}\\
		&=\begin{bmatrix}
			L_1\quad 1\\
			L_2\quad 1\\
			...\quad...\\
			L_m\quad 1
		\end{bmatrix} [\sigma^2_\text{s} \quad \sigma^2_\text{r}]^\text{T}
	\end{aligned}
	\label{eq:LoNPE}
\end{equation}
if only $\textit{Rank}([\mathbf{L},\ \mathbf{1}])\ge2$ where $\mathbf{L}=[L_1, L_2, ..., L_m]^\text{T}$, it would be effective to estimate the noise prior $(\sigma_\text{s},\ \sigma_\text{r})$ using a simple least square optimization algorithm.

Note that, LoNPE algorithm estimate the noise prior from a single raw noisy observation $\mathbf{I}$ (or $\mathbf{I}_\text{raw}$) as
\begin{equation}
	[\sigma_\text{s}, \sigma_\text{r}] = \Psi_\text{LoNPE}(\mathbf{I}_\text{raw})
	\label{eq:LoNPE_algorithm}
\end{equation}
which strictly follows the statistical characteristics of raw sensor noise prior, it might be infeasible to directly apply to a sRGB color image due to the sophisticated ISP procedure. 

\begin{figure}[!t]
	\centering
	\includegraphics[width=1\linewidth]{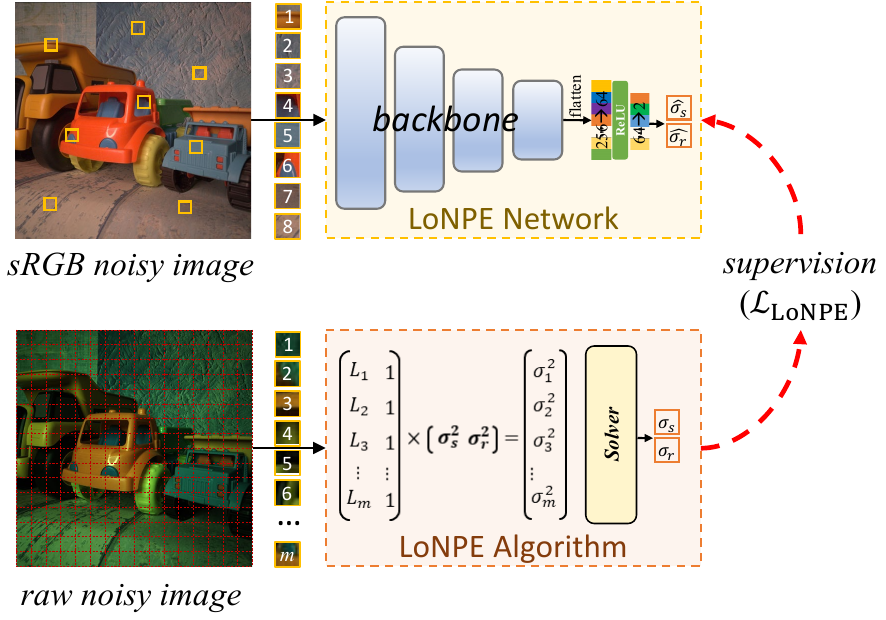}
	\caption{Framework of LoNPE network. The backbone of network is same as the one in DoTNet \citep{HuangY2023TPAMI}.}
	\label{fig:LoNPE Network}
\end{figure}
\subsubsection{LoNPE Network}\label{sec:LoNPE Network}
Inspired by the concept of noise representation in \citet{HuangY2023TPAMI}, that the noise level could be represented as an explicit parameter and is easy to learn by an external neural network. We build a learnable CNN model to predict the noise prior $(\hat\sigma_\text{s},\hat\sigma_\text{r})\rightarrow(\sigma_\text{s}, \sigma_\text{r})$ from a single sRGB noisy color image $\mathbf{I}_\text{rgb}$, which is formulated as 
\begin{equation}
	[\hat \sigma_\text{s}, \hat \sigma_\text{r}] = \Phi_\text{LoNPE}(\mathbf{I}_\text{rgb})
	\label{eq:LoNPE_model}
\end{equation}
where $\Phi_\text{LoNPE}(\cdot)$ denotes the noise prior estimation network, and called ``LoNPE network''. 
The framework of LoNPE network is shown in Fig.~\ref{fig:LoNPE Network}, consisting of a backbone as the DoTNet~\citep{HuangY2023TPAMI} for feature extraction and two fully-connected layers for decision. Specifically, we sample only 8 random local patches from each image to estimate an accurate noise prior, instead of using $m$ patches as previously described. Due to the physical characteristics of shot noise (photon-to-electron conversion rate) and read noise (quantization range of digital signal), the output range of $\sigma_\text{s}$ and $\sigma_\text{r}$ are limited to $[0,1]$.

To learn an effective LoNPE network, we need to calculate the groundtruth noise prior, by applying the LoNPE algorithm $\Phi_\text{LoNPE}(\cdot)$ with numerous training samples of raw noisy observations, \eg, SIDD-Medium raw-domain dataset. Subsequently, we train the LoNPE network by optimizing the following objective function,
\begin{equation}
	\mathcal{L}_\text{LoNPE} = \left \| \Phi_\text{LoNPE}(\mathbf{I}_\text{rgb}) - \Psi_\text{LoNPE}(\mathbf{I}_\text{raw}) \right \| _1
	\label{eq:objective_LoNPE}
\end{equation}
where $\left\|\cdot \right\|_1$ represents L1 loss function, minimizing the mean absolute error (MAE) between the estimated noise prior and the corresponding groundtruth.

\begin{figure*}[!t]
	\centering
	\includegraphics[width=1\linewidth]{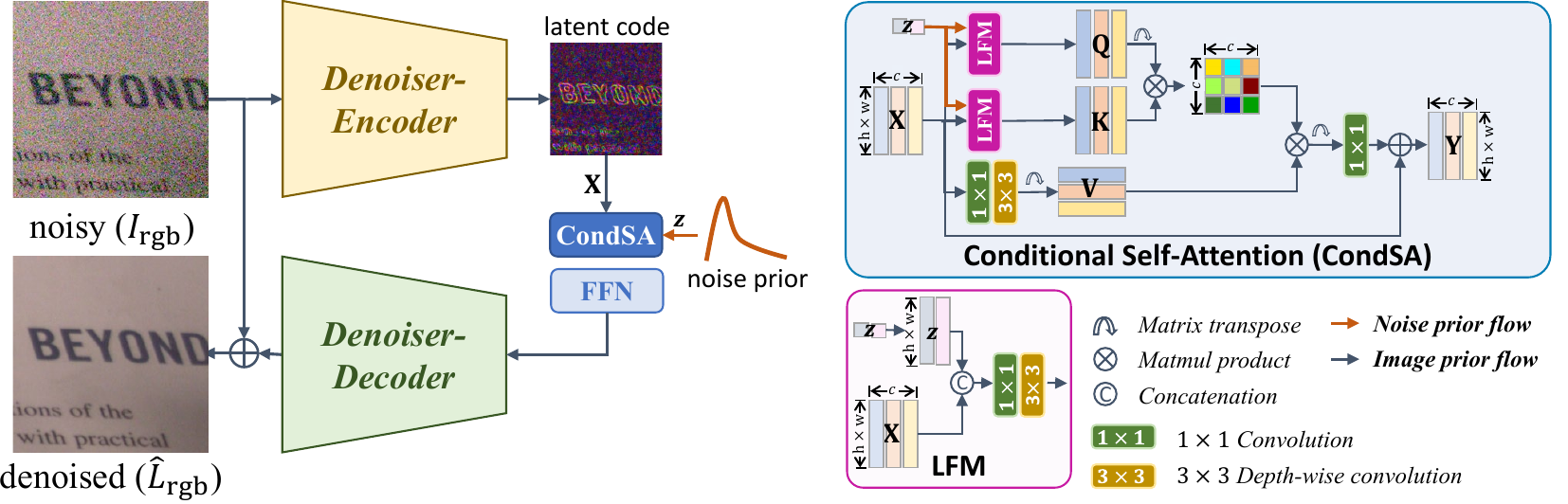}
	\caption{Architecture of Condformer. The information flow is successively transmitted into the denoiser encoder, the latent module, and the denoiser decoder, where the latent module stacks groups of CondSA blocks. Particularly, the noise prior is generated via LoNPE network, and is then embedded as a conditional vector $\mathbf{z}$ into the query tensor $\mathbf{Q}$ and key tensor $\mathbf{K}$ in each CondSA block.}
	\label{fig:Condformer}
\end{figure*}
\subsection{Conditional Denoising Transformer}\label{sec:Condformer}
As discussed in Section~\ref{sec1}, a conditional denoising model is necessary for improving the performance by decomposing the optimization space under the guidance of noise prior. Considering the image prior and noise prior in a noisy image, an excellent conditional denoising model should be good at extracting the implicit image prior to learn how to restore the corrupted pixels, and utilizing the explicit noise prior to precisely control the intensity of restoration. 

Due to the strong capability of Transformer-based denoising models (\eg, Uformer \citep{WangZ2022CVPR}, Restormer \citep{ZamirS2022CVPR}, GRL~\cite{LiY2023CVPR} and \etc), we design a Conditional denoising Transformer ({\bf Condformer}) by embedding the noise prior into the self-attention module. Particularly, as the independence of noise prior and image prior, guiding the model to learn from image prior and noise prior separately is the primary principle in designing the Condformer. 

\subsubsection{Embedding noise prior in latent space}
Existing image denoising networks typically use a global residual connection for noise prediction and an U-shape encoder-decoder structure for feature representation, suggesting that noise is implicitly concealed in the latent space as illustrated in Fig.~\ref{fig:Condformer}. Meanwhile, image prior such as scene context is theoretically weakest in the latent space. Inspired by this and considering the residual attribute of noise, the latent space code of a noisy image significantly represents noise statistics.

As mentioned earlier in Section~\ref{sec1}, an effective conditional denoising model should separately consider the image and noise priors, adhering to the principle of their independence. Therefore, the noise prior should be embedded in the latent space to strengthen noise statistics representation and guide the denoising network to focus more on noisy residues.
Specifically as shown in Fig.~\ref{fig:Condformer}, we firstly extract the latent space code $\mathbf{X}$ of the noisy image using a denoiser encoder, and then construct a feature fusion module to embed the noise prior $(\sigma_\text{s}, \sigma_\text{r})$.

\subsubsection{Overall pipeline}
Based on the principle of embedding the noise prior in the latent space, it is feasible to incorporate the noise prior into any encoder-decoder denoiser. Considering the efficiency in practical applications, we employ the Restormer as our denoiser baseline, replacing its self-attention module by a conditional self-attention module that embeds noise prior in the latent space, while keeping all other modules unchanged. Following \citet{ZamirS2022CVPR}, given a noisy color observation $\mathbf{I}_\text{rgb}\in\mathbb{R}^{3\times h\times w}$, we construct a multi-scale hierarchical denoiser encoder which consists of three channel-wise Transformer blocks to capture cross-covariance across channels, generating the latent space code $\mathbf{X}\in\mathbb{R}^{8c\times\frac{h}{8}\times\frac{w}{8}}$, where $c$ is the number of channels. 

On handling the latent space code, we introduce a conditional self-attention module (CondSA) that embeds the noise prior into the implicit latent code for conditional optimization of denoising model. The rectified latent code is 
\begin{equation}
	{\mathbf{Y}}=\text{CondSA}(\mathbf{X}, \mathbf{z})
\end{equation}
This is then fed into a feed-forward network (FFN) for feature transformation. In particular, to effectively exploit the noise prior $(\sigma_\text{s}, \sigma_\text{r})$ estimated by LoNPE algorithm or network, this prior is firstly encoded into a latent conditional embedding vector $\mathbf{z}\in\mathbb{R}^{1\times c_z}$ using a shallow module with fully-connected layers. Indicating single CondSA has a specific embedding vector to adapt the intermediate features. 
The rectified latent space code equipped with noise prior related embedding, guides the denoiser decoder for specific denoising. In particularly, to preserve fine structural and textural details in the restored images, we use a hierarchical skip-connection strategy \citep{ZamirS2022CVPR} that integrates low-level features from the encoder and high-level features from the decoder.

Consequently, as depicted in Fig.~\ref{fig:Condformer}, the denoised output $\hat{\mathbf{L}}_\text{rgb}$ can be formulated as 
\begin{equation}
	\hat{\mathbf{L}}_\text{rgb} = \Phi_\text{Condformer}(\mathbf{I}_\text{rgb}, (\sigma_\text{s}, \sigma_\text{r}))
\end{equation}
where $\Phi_\text{Condformer}(\cdot)$ represents the Condformer model, with inputs consisting of a noisy observation $\mathbf{I}_\text{rgb}$ and the estimated noise prior $(\sigma_\text{s}, \sigma_\text{r})$ from LoNPE algorithm $\Phi_\text{LoNPE}(\mathbf{I}_\text{raw})$ or network $\Psi_\text{LoNPE}(\mathbf{I}_\text{rgb})$. 

Following the convention of supervised denoising methods, we train the Condformer by optimizing the pixel-wise objective function as
\begin{equation}
	\mathcal{L}_\text{Condformer} = \left \| \mathbf{L}_\text{rgb} - \hat{\mathbf{L}}_\text{rgb} \right \| _1
\end{equation}
where $\mathbf{L}_\text{rgb}$ denotes the clean target corresponding to the denoised output.

Subsequently, we introduce the preliminary definition of the self-attention module in Restormer, and describe the proposed CondSA for embedding the conditional noise prior.

\subsubsection{Conditional Self-Attention Mechanism}
Aiming to alleviate the high complexity of the conventional self-attention module when calculating the key-query cross-covariance across spatial dimensions, the self-attention in Restormer attempts to calculate the key-query cross-covariance across channels, and is formulated as
\begin{equation}
	\mathbf{Y} = W^Y \text{Attention}(\mathbf{Q}, \mathbf{K}, \mathbf{V})+\mathbf{X}
\end{equation}
that,
\begin{equation}
	\text{Attention}(\mathbf{Q}, \mathbf{K}, \mathbf{V}) = \text{Softmax}({\mathbf{Q}\mathbf{K}^\textsf{T}}/{\alpha})\mathbf{V}
\end{equation}
where $\mathbf{X}$ and $\mathbf{Y}$ are the input and output features. $\mathbf{Q}\in\mathbb{R}^{c\times hw}$, $\mathbf{K}\in\mathbb{R}^{c\times hw}$ and $\mathbf{V}\in\mathbb{R}^{c\times hw}$ indicate the {\em query}, {\em key} and {\em value} matrix obtained by encoding the input feature $\mathbf{X}\in\mathbb{R}^{c\times h\times w}$ with linear layers $W^Q$, $W^K$ and $W^V$ respectively, each of which stacks a $1\times1$ convolution and a $3\times3$ depth-wise convolution layer. Besides, $\alpha$ is a learnable scaling parameter to control the magnitude of the cross-covariance of $\mathbf{Q}$ and $\mathbf{K}$ before applying a softmax layer.

By extracting and exploring the local and non-local features, the whole network indeed aims to employ the image priors of the noisy observation, which exactly meets the unconditional optimization paradigm in Fig.~\ref{fig:optim_a}. According to the optimization of conditional denoising model in Eq.(\ref{eq:cond_optim}) and Fig.~\ref{fig:optim_b}, the noise prior should be embedded into a conditional self-attention module as
\begin{equation}
	\mathbf{Y} = W^Y \text{Attention}(\mathbf{Q}, \mathbf{K}, \mathbf{V}, \mathbf{z})+\mathbf{X}
\end{equation}
where $\mathbf{z}\in\mathbb{R}^{1\times 2c_z}$ represents the conditional embedding vector from the noise prior $(\sigma_\text{s}, \sigma_\text{r})$ by repeating $k$ times in channel dimension. 

The conditional attention should effectively represent the correlation between-in the intermediate image features, and the latent correlation between the intermediate image features and noise prior. Therefore, as mentioned in Section 3.1.1, a feature fusion module is essential for capturing the relationship between the noise prior and latent code. Intuitively, the query tensor $\mathbf{Q}$ and the key tensor $\mathbf{K}$ indicates the information for feature retrieval \citep{VaswaniA2017NeurIPS}; instead, the value tensor $\mathbf{V}$ represent the property of the input feature. Therefore, it is reasonable to embed noise prior into the query/key tensors and generate the conditional counterparts $\mathbf{Q}'$ and $\mathbf{K}'$, so we have
\begin{equation}
	\text{Attention}(\mathbf{Q}, \mathbf{K}, \mathbf{V}, \mathbf{z}) = \text{Softmax}(\mathbf{Q}'{\mathbf{K}'}^\textsf{T}/{\alpha})\mathbf{V}
\end{equation}\vspace{-0.1cm}
and 
\begin{equation}
	\mathbf{Q}' = {\Phi}_\text{LFM}(\mathbf{Q}, \mathbf{z}),\quad \mathbf{K}' = {\Phi}_\text{LFM}(\mathbf{K}, \mathbf{z})
\end{equation}
where $\Phi_\text{LFM}$ represents a linear fusion module (LFM) layer for feature fusion. In particular, since noise is distributed across both spatial and channel dimensions of intermediate image features, the LFM fuses the query/key tensors and the conditional embedding vector in both dimensions by concatenating them along the channel axis and applying a $1\times1$ convolution and a $3\times3$ depth-wise convolution layer. This design enables effective and localized integration of noise prior into the feature representations.

\section{Experiments}~\label{sec:experiments}
We describe the experimental setup and then evaluate the performances of our LoNPE and Condformer on noise statistics and blind image denoising. Finally, we perform ablation studies to demonstrate the effectiveness of our methods.

\subsection{Experimental Setup}
\subsubsection{Datasets and Metrics} \label{sec: datasets}
To demonstrate the effectiveness of our LoNPE and Condformer, we conduct experiments on both synthetic and real datasets. Below are the details of the synthetic and real datasets, and evaluation metrics.

\noindent{\bf \em Synthetic Datasets}. Following \citep{ZhangK2021TPAMI}, we adopt several sRGB image datasets for training the LoNPE and Condformer networks, including the DIV2K and Flicker2K dataset \citep{AgustssonE2017CVPRW} with 3650 high-quality 2K images, the Berkeley segmentation dataset (BSD) \citep{MartinF2001ICCV} with 400 images, Waterloo Exploration Database (WED) dataset \citep{MaK2016TIP} with 4744 images. Considering both signal-dependent and signal-independent noises, we randomly add noises on the clean sRGB image with Poisson-Gaussian noise model. The noise level are set to $\sigma_\text{s}\sim\mathbb{U}(0, 0.3)$ and $\sigma_\text{r}\sim\mathbb{U}(0, 50/255)$, both of which indicate the groundtruth noise prior to demonstrate the effectiveness of our LoNPE algorithm. Especially, the Poisson-Gaussian noise will be degraded to additive Gaussian white noise (AWGN) when $\sigma_\text{s}=0$. To evaluate the performance, we apply our methods on several benchmarks, including CBSD68 \citep{MartinF2001ICCV}, Kodak24 \citep{FranzenR1999Kodak} and Urban100 \citep{HuangJ2015CVPR}.

\noindent{\bf \em Real Datasets}. Consistent with previous real image denoising work \citep{ZamirS2022CVPR}, we adopt the SIDD-Medium dataset \citep{AbdelhamedA2018CVPR} for real noise statistics and real image denoising tasks, which contains 320 noisy-clean image pairs in both raw and sRGB domains. 
Particularly, we first apply our LoNPE algorithm on 320 noisy raw images to estimate their corresponding noise priors, and then utilize them to help training the LoNPE and Condformer networks on the 320 sRGB noisy-clean image pairs. For validation, 1024 pairs of noisy-clean sRGB image patches from SIDD validation dataset \citep{AbdelhamedA2018CVPR} are adopted. Besides, evaluation is also conducted on 1280 noisy $256\times256$ patches from the SIDD benchmark dataset \citep{AbdelhamedA2018CVPR} and 50 noisy $512\times512$ images from the DND benchmark dataset \citep{PlotzT2017CVPR}. 

\noindent{\bf \em Evaluation Metrics}. Two commonly-used image quality assessment criteria are adopted to evaluate the performances: Peak Signal-to-Noise Ratio (PSNR) and Structural SIMilarity (SSIM) \citep{WangZ2004TIP}, and are calculated in the sRGB domain. Note that, since the clean groundtruthes of SIDD and DND benchmark datasets are unavailable, we calculate these metrics on only SIDD validation dataset and obtain the metrics of them from the online servers\footnote{SIDD: \href{https://abdokamel.github.io/sidd/}{https://abdokamel.github.io/sidd/}; \\ DND: \href{https://noise.visinf.tu-darmstadt.de}{https://noise.visinf.tu-darmstadt.de}}.

\begin{sidewaystable}
	\centering
	\small
	\renewcommand\arraystretch{1.1}
	\caption{Comparative study of noise prior estimation on the Urban100 dataset. Random noises are sampled based on the given noise prior parameters $(\sigma_\text{s}, \sigma_\text{r})$ and added to each clean image. Note that the top three methods are executed on a CPU platform, while the bottom three methods are executed on a GPU platform.}
		\begin{tabular}{p{7cm} | c | c | c c c| r}
			\hline
			\hline
			\multirow{3}{*}{Methods}&\multicolumn{5}{c|}{Noise prior parameter estimation result of $(\sigma_\text{s}, \sigma_\text{r})$}&\multirow{3}{*}{\tabincell{c}{Time\\(s)}}\\
			&Gaussian&Poisson&\multicolumn{3}{c|}{Poisson-Gaussian}&\\
			&(\textcolor{gray}{0.000}, 0.050)&(0.100, \textcolor{gray}{0.000})&(0.050, 0.020)&(0.100, 0.050)&(0.200, 0.100)&\\
			\hline
			\citet{MakitaloM2014TIP}&(\textcolor{gray}{0.441}, \textbf{0.044})&(0.111, \textcolor{gray}{0.006})&(0.186, 0.018)&(0.446, 0.041)&(0.866, 0.089)&56.32\\
			\citet{PimpalkhuteV2021TIP}&(\textcolor{gray}{\quad-\quad}, 0.039)&(\quad-\quad, \quad-\quad)&(\quad-\quad, \textbf{0.021})&(\quad-\quad, 0.039)&(\quad-\quad, 0.075)&\textbf{0.08}\\
			LoNPE Algorithm (Ours)&(\textcolor{gray}{0.053}, 0.066)&(\textbf{0.104}, \textcolor{gray}{0.009})&(\textbf{0.062}, 0.029)&(\textbf{0.117}, \textbf{0.057})&(\textbf{0.233}, \textbf{0.091})&0.17\\
			\hline
			FBI-Denoiser$_\text{(0.05,0.02)}$~\citep{ByunJ2021CVPR}&(\quad-\quad, \quad-\quad)&(\quad-\quad, \quad-\quad)&(0.041,0.004)&(\quad-\quad, \quad-\quad)&(\quad-\quad, \quad-\quad)&0.08\\
			FBI-Denoiser$_\text{(mixed)}$~\citep{ByunJ2021CVPR}&(\quad-\quad, \quad-\quad)&(\quad-\quad, \quad-\quad)&(0.010,0.012)&(0.017, 0.021)&(0.089, 0.136)&0.08\\
			LoNPE Network (Ours)&(\textcolor{gray}{0.022}, \textbf{0.043})&(\textbf{0.088}, \textcolor{gray}{0.015})&(\textbf{0.046}, \textbf{0.020})&(\textbf{0.092}, \textbf{0.045})&(\textbf{0.182}, \textbf{0.075})&\textbf{0.01}\\
			\hline
			\hline
		\end{tabular}
		\label{tab:noise estimation}
	\end{sidewaystable}

\begin{figure*}
	\centering
	\includegraphics[width=1.\linewidth]{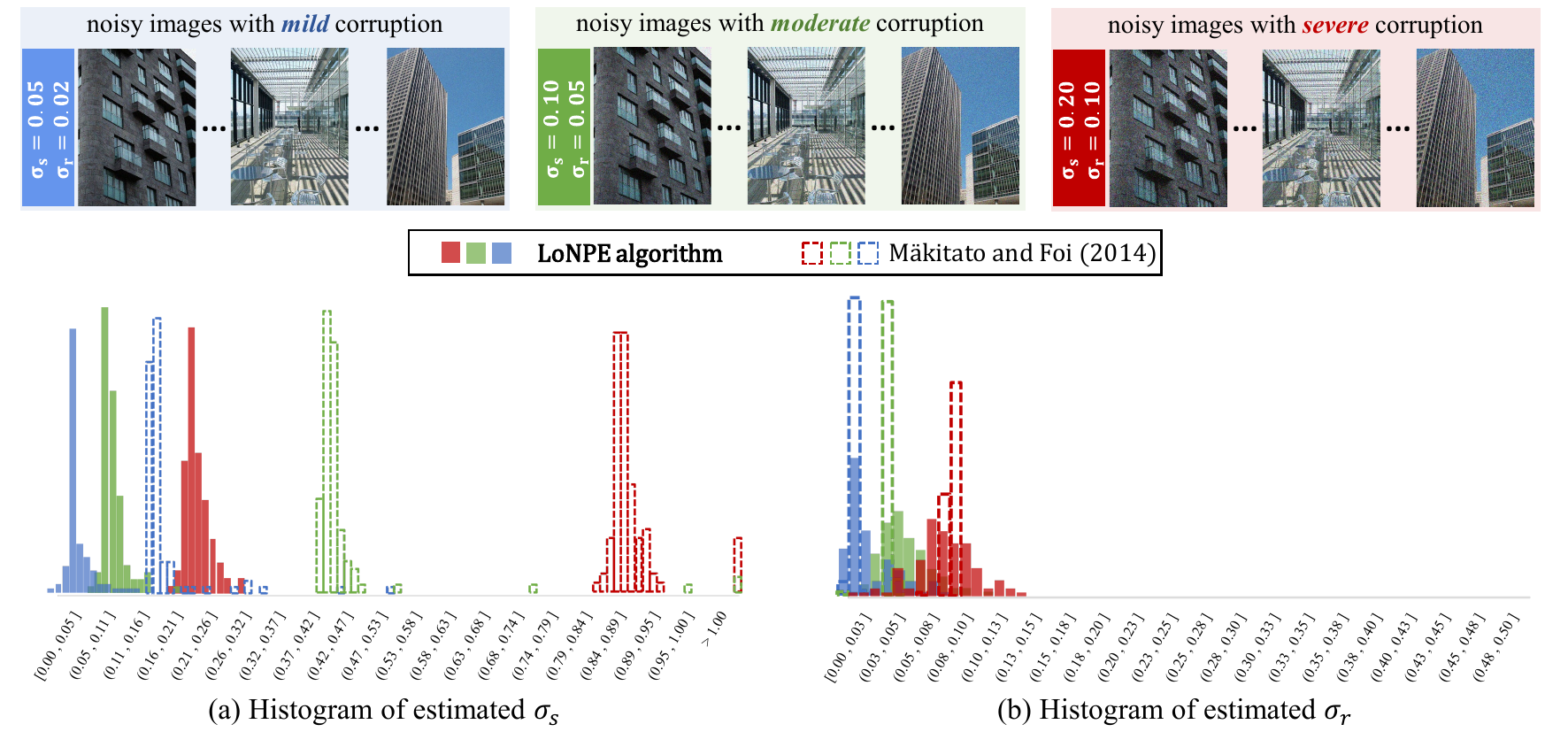}
	\caption{Statistical study on noise prior estimation performance of \cite{MakitaloM2014TIP} and our LoNPE algorithm with synthetic Poisson-Gaussian noises. By applying our LoNPE algorithm, we can effectively calculate a relative accurate noise prior from a single noisy observation.}
	\label{fig:noise estimation on P-G distribution}
\end{figure*}
\subsubsection{Implementation Details} 
\noindent{\bf\em Settings on LoNPE Algorithm}. As described in Section~\ref{sec:LoNPE Algorithm}, to accurately calculate the noise prior of a noisy observation in raw domain, our LoNPE algorithm $\Psi_\text{LoNPE}$ is applied on top $\frac{m}{n}=10\%$ of the sampled local patches, which are at size of $\mathcal{O}=16\times 16$ for each raw image. Given a $h \times w$ raw noisy image and set the sampling stride to be $k=4$, we first sample $n=\left \lfloor \frac{h}{k} \right \rfloor \times \left \lfloor \frac{w}{k} \right \rfloor$ local patches, and select the top $m$ smooth patches with lower $\lambda_\mathcal{S}$, then calculate its noise prior using Eq.(\ref{eq:LoNPE_algorithm}).

\noindent{\bf\em Settings on LoNPE Network}. As described in Section~\ref{sec:LoNPE Network}, for practical applications on estimating the noise prior of a sRGB noisy image, we need to train an effective LoNPE network $\Phi_\text{LoNPE}$ as illustrated in Fig.~\ref{fig:LoNPE Network}. Similar to \citep{HuangY2023TPAMI}, for a single noisy image, we randomly sample 8 local patches of size $32\times32$ and average the output $(\hat\sigma_\text{s},\hat\sigma_\text{r})$ as the final predicted noise prior. In the training phase, AdamW optimizer \citep{LoshchilovI2019ICLR} is employed with cosine annealing \citep{LoshchilovI2017ICLR} learning rate from $10^{-3}$ to $10^{-6}$ during 50K mini-batch iterations, on minimizing the objective function in Eq.(\ref{eq:objective_LoNPE}), and the batch size is set to 64.

\noindent{\bf\em Settings on Condformer}. Following the settings of Restormer in \citep{ZamirS2022CVPR}, we build our Condformer with groups of Transformer block in the encoder or decoder modules. Yet in the latent module, we stack 8 CondSA blocks with the predicted noise prior to recify the latent space. In each CondSA block, the length of embedding vector $\mathbf{z}$ is set to $c_z=c=48$, where $c$ denotes the initial feature channels of encoder and indicates the latent feature has $8c=384$ channels.
In the training phase, we adopt the AdamW optimizer with $(\beta_1, \beta_2)=(0.9, 0.999)$ and set weight decay to $10^{-4}$. Using progressive training strategy proposed by \citep{ZamirS2022CVPR}, we set the batch size and patch size pairs to [$(64, 128^2)$, $(16, 256^2)$, $(8, 384^2)$, $(4, 512^2)$] at training iterations [0k, 150k, 200k, 250k]. We train our Condformer models for total 300k iterations and the initial learning rate is set to $4\times10^{-4}$ and gradually reduced to $10^{-6}$ through the cosine annealing. Data augmentation is performed on the training data through horizontal flip and random rotation of 90, 180, and 270.

Both of the LoNPE and Condformer networks are implemented on PyTorch framework using NVIDIA A800 GPUs.

\begin{figure*}
	\centering
	\includegraphics[width=1\linewidth]{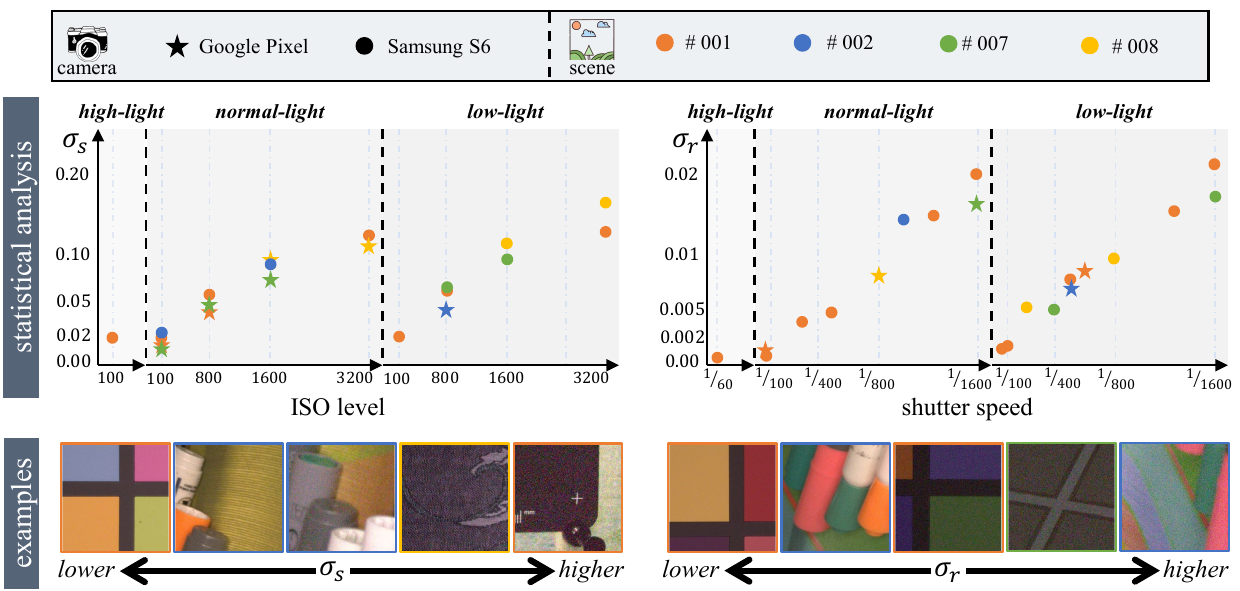}
	\caption{Statistical experiments on noise prior of raw sensor images from various scenes in SIDD-Medium training datasets. We can find that different imaging environments cause various noise priors, \eg, higher ISO level causes more shot noises and higher shutter speed causes more read noises. Instead, different target scenes show similar noise priors, indicating the independence of image prior and noise prior.}
	\label{fig:Statistical analysis on SIDD}
\end{figure*}
\subsection{Experiments on Noise Statistics}
In this section, we conduct several statistical experiments to verify the effectiveness of our LoNPE algorithm and network on noise prior representation. In particular, we first conduct noise prior estimation experiment on synthetic Poisson-Gaussian noises quantitatively, and further analyze the statistics of real noise. 

\subsubsection{On synthetic noises}\label{sec:synthetic statistical study}
As illustrated in Section~\ref{sec:independency}, the noise prior is beyond and independent of the image prior. From Eq. (\ref{eq: digital raw signal}), the noise prior $(\sigma_\text{s}, \sigma_\text{r})$ affects raw noisy observations but is implicit in the sRGB color observations due to unknown ISP operations. To address this limitation, we randomly add noises to clean sRGB images with Poisson-Gaussian sampling and employ our LoNPE algorithm on the synthesized noisy images to estimate the corresponding noise prior. 

Quantitatively, we perform a comparative study on noise prior parameter estimation, as reported in Table~\ref{tab:noise estimation}. By setting various noise prior parameters, Gaussian, Poisson and the complicated Poisson-Gaussian noise types are considered. For the comparison of synthetic noise prior estimation, several noise parameter estimation methods are evaluated, including the traditional method for Gaussian \citep{PimpalkhuteV2021TIP} or Poisson-Gaussian noises \citep{MakitaloM2014TIP}, and the CNN-based Poisson-Gaussian noise parameter estimators in FBI-Denoiser~\citep{ByunJ2021CVPR} with fixed and mixed noise levels.	
Particularly for Poisson-Gaussian noises, it is observed that \citet{MakitaloM2014TIP} fails when signal-independent noise is more severe than signal-dependent noise, and commonly requires significant computation time to search for intersection of the unitary variance contours. In contrast, our LoNPE algorithm effectively handles more general scenarios, including Gaussian, Poisson and Gaussian-Poisson noises, with a speedup of up to $\times300$. Additionally, by leveraging GPU acceleration, our LoNPE network increases speed significantly with negligible performance degradation then LoNPE algorithm, and achieves superior generalization across all scenarios compared to the noise parameter estimator in FBI-Denoiser~\citep{ByunJ2021CVPR}. 

To further demonstrate the stability of our LoNPE algorithm, we visualize the statistical histogram of the estimated noise priors. As shown in Fig.~\ref{fig:noise estimation on P-G distribution}, we set three levels of noise prior parameter $(\sigma_\text{s}, \sigma_\text{r})$ as $(0.05, 0.02)$, $(0.10, 0.05)$ and $(0.2, 0.1)$, which is consistent with the setting in Table~\ref{tab:noise estimation}, representing mild, moderate and severe corruptions, respectively. By applying LoNPE algorithm, we obtain the estimated noise prior parameters of each image with different noise level, and find that the mean values of estimation results are closer to the groundtruthes than the compared method \citet{MakitaloM2014TIP}, particularly for the signal-dependent noise prior $\sigma_\text{s}$. Nevertheless, there exists a nonnegligible discrepancy in the signal-independent noise prior $\sigma_\text{r}$ estimation, and the discrepancy increases as the noise level is enlarged. A reason is that, the image context reflects the signal intensity of the target scene illuminance, and could interfere the estimation of the signal-independent $\sigma_\text{r}$. This issue is particularly pronounced at the high-frequency regions, \eg, edges and textures. That is why we need to employ the local smoothness criterion $\lambda_\mathcal{S}$ to eliminate the interference of image prior. 

\subsubsection{On real noises}
Furthermore, as described in Section~\ref{sec:LoNPE}, the goal of LoNPE is to calculate the noise prior from a single raw noisy observation based on the independence of noise prior and image prior. Thus, the most critical effectiveness demonstration for real scenes is a statistical experiment on the correlations of the estimated noise prior $(\sigma_\text{s}, \sigma_\text{r})$ and the camera sensor imaging environments. 

Based on the independence of image prior and noise prior investigated in Fig.~\ref{fig:Noise Prior beyond Image} and the statistical study on synthetic Poisson-Gaussian noise prior estimation in Section~\ref{sec:synthetic statistical study}, the noise prior is quantifiable using our LoNPE algorithm. Thus, we conduct a statistical study on the estimated noise priors of SIDD noisy observations with various target scenes and imaging environments. Specifically, according to the camera information of the SIDD-Medium raw-domain training dataset~\citep{AbdelhamedA2018CVPR}, we mainly analyze the estimated noise prior under various brightness (including ``high-light'', ``normal-light'' and ``low-light''), scenes (including scene ``001'', ``002'', ``007'' and ``008''), ISO levels (ranging from 100 to 3200), and shutter speed (ranging from 1/1600 to 1/60). Subsequently, the statistical results in Fig.~\ref{fig:Statistical analysis on SIDD} can be summarized into several points:

\noindent{\bf \em 1) Higher ISO level leads to more shot noise.} Shot noise prior $\sigma_\text{s}$ grows with ISO levels, as higher ISO amplifies the signal generated by photons on the camera sensor. Since shot noise follows a Poisson distribution, amplification makes the noise more pronounced, especially in low-light conditions, affecting image quality. Thus, the slope of variance ($\sigma^2_\text{s}$) in Eq.~(\ref{eq: digital raw signal}) would be larger as the ISO increased.

\noindent{\bf \em 2) Higher shutter speed leads to more read noises.} Under same brightness, the read noise prior $\sigma_\text{r}$ scales with shutter speed. Faster shutter speeds, particularly in high-speed photography, require rapid sensor readout, which introduces additional electronic noise, leading to higher read noise.

\noindent{\bf \em 3) Independence of noise and image priors.} Brightness influences shot noise via photon intensity, while ISO and shutter speed affect shot and read noise, respectively, without altering the scene-derived signal. Statistical analysis confirms that noise characteristics remain consistent across different scenes under the same imaging conditions, proving the independence of noise priors from image priors.

\begin{sidewaystable}
	\renewcommand\arraystretch{1.}
	\small
	\centering
	\caption{Quantitative comparisons of different synthetic image denoising methods on several validation datasets with a fixed $\sigma_\text{s}$ and various $\sigma_\text{r}\in[15, 25, 50]/255$. PSNR$\uparrow$ criterion is adopted to evaluate the performances.}
	\begin{tabular}{l c c| c c c | c c c | c c c}
		\hline
		\hline
		\multirow{2}{*}{Method}&{Noise}&Params&\multicolumn{3}{c|}{CBSD68}&\multicolumn{3}{c|}{Kodak24}&\multicolumn{3}{c}{Urban100}\\
		&Model&(M)&$\sigma_\text{s}$=0&$\sigma_\text{s}$=0.15&$\sigma_\text{s}$=0.3&$\sigma_\text{s}$=0&$\sigma_\text{s}$=0.15&$\sigma_\text{s}$=0.3&$\sigma_\text{s}$=0&$\sigma_\text{s}$=0.15&$\sigma_\text{s}$=0.3\\
		\hline
		Noisy&-&-&20.12&17.10&13.88&20.02&16.97&13.72&20.27&17.12&13.87\\
		DnCNN \citep{ZhangK2017TIP}&G&0.7&31.00&29.10&26.30&31.89&30.12&27.27&30.46&28.70&25.64\\
		Restormer \citep{ZamirS2022CVPR}&G&26.1&\underline{31.57}&28.98&25.40&\textbf{32.82}&30.24&26.62&\textbf{32.65}&29.84&25.68\\
		\hline
		VDN \citep{YueZ2019NeurIPS}&SV-G&2.0&31.21&29.33&27.22&32.28&30.49&28.33&31.46&29.61&27.05\\
		DRANet \citep{WuW2024PR}&SV-G&1.6&31.33&29.44&27.30&32.50&30.69&28.44&31.93&30.09&27.65\\
		VIRNet \citep{YueZ2024TPAMI}&SV-G&10.5&31.45&29.52&27.17&32.65&30.76&28.24&32.23&30.24&27.14\\
		\hline
		FBI-Denoiser \citep{ByunJ2021CVPR}&P-G&7.5&26.59&25.55&23.30&27.84&26.69&24.11&25.60&24.56&22.32\\
		VBDNet \citep{LiangH2023PR}&P-G&8.3&25.93&25.51&24.41&27.00&26.73&25.68&26.45&25.86&24.57\\
		CLIPDenoising \citep{ChengJ2024CVPR}&P-G&34.5&30.52&28.18&25.19&31.31&29.02&25.92&30.04&27.79&24.48\\
		\hdashline
		Restormer$^\dagger$ \citep{ZamirS2022CVPR}&P-G&26.1&31.48&\underline{29.66}&\underline{28.01}&\underline{32.70}&\underline{31.02}&\underline{29.44}&32.46&\underline{30.95}&\underline{29.37}\\
		MambaIR$^\dagger$ \citep{GuoH2024ECCV}&P-G&23.2&31.41&29.60&27.94&32.65&30.99&29.41&32.42&30.91&29.32\\
		Condformer (Ours)&P-G&27.0&\textbf{31.59}&\textbf{29.84}&\textbf{28.14}&\underline{32.81}&\textbf{31.20}&\textbf{29.58}&\underline{32.64}&\textbf{31.15}&\textbf{29.56}\\
		\hline
		\hline
		\multicolumn{11}{l}{\footnotesize$^\dagger$: Model is re-trained on Poisson-Gaussian noise model with the same settings as our Condformer.}
	\end{tabular}
	\label{tab:SOTA_Poisson-Gaussian}
\end{sidewaystable}

\begin{figure*}[!t]
	\centering
	\includegraphics[width=1\linewidth]{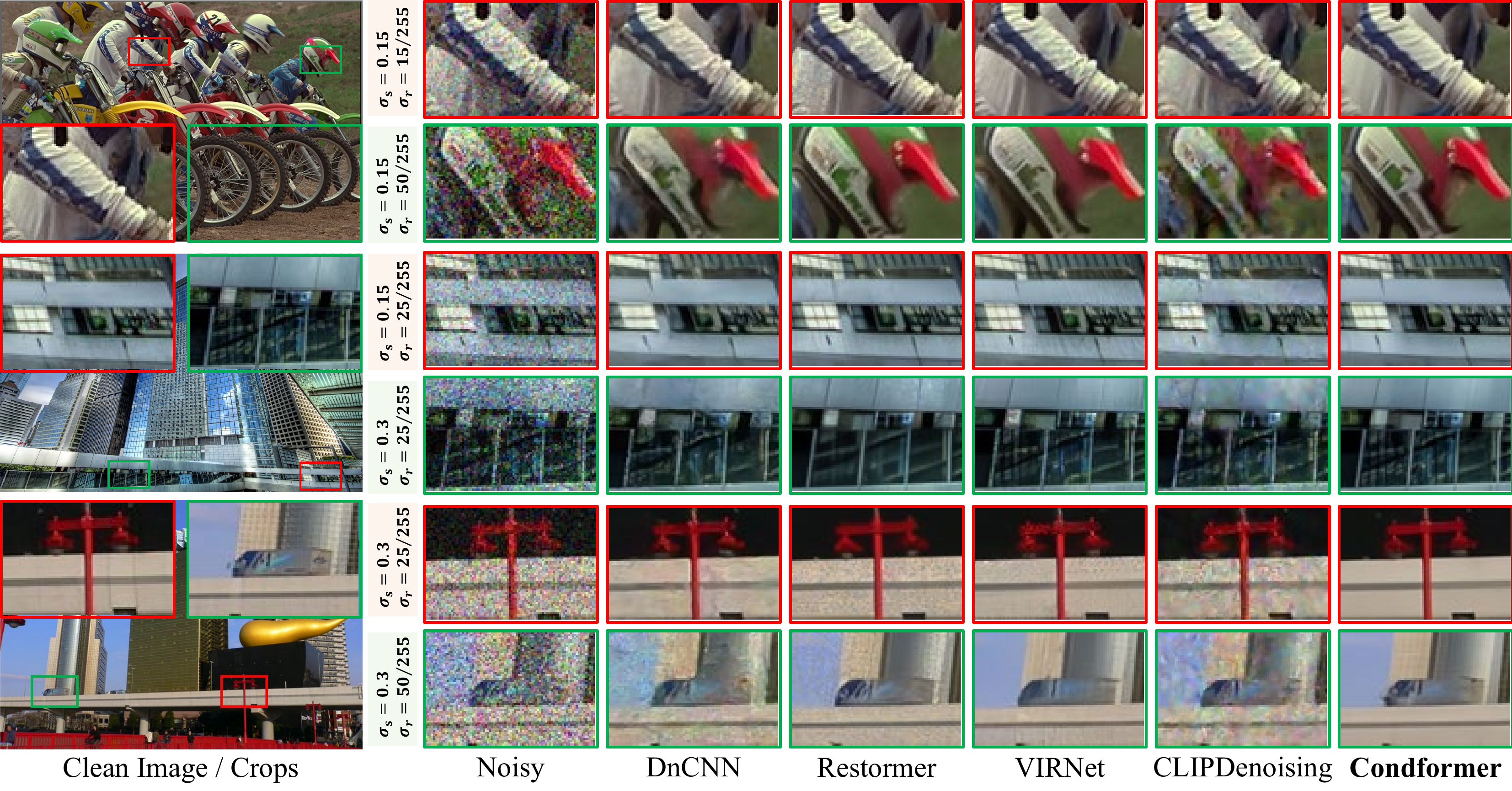}
	\caption{Visual results of restoring the Poisson-Gaussian noisy images. We can find that our Condformer can preserve the details and remain less distortions as compared with other methods, showing higher generalization on various corruptions with different noise levels.}
	\label{fig:visual results_Poisson-Gaussian}
\end{figure*}
Consequently, understanding the independence of noise prior and image prior allows photographers and engineers to develop better noise removal algorithms and improve sensor designs. By treating the scene and sensor noise as separate entities, it becomes easier to process images to enhance the desired signal intensity while minimizing the impact of sensor noise on the final image.


\subsection{Experiments on Image Denoising}
In this section, we mainly conduct quantitatively and qualitatively experimental study on synthetic and real blind image denoising performance of the proposed Condformer. 

\subsubsection{On synthetic images}
Due to the agnosticism of real noise prior, it is hard to conduct a comprehensive validation on image denoising under various noise priors. We firstly conduct Poisson-Gaussian blind image denoising experiments on synthetic validation datasets. We mainly synthesize three levels of shot noise prior $\sigma_\text{s}\in[0, 0.15, 0.3]$, and sample read noise priors $\sigma_\text{r}\in[15,25,50]/255$ with each fixed $\sigma_\text{s}$. Particularly, this setting follows the convention of mainstream Gaussian blind image denoising researches as $\sigma_\text{s}=0$ and $\sigma_\text{r}\in[15,25,50]/255$.

Recent blind denoisers were commonly trained on noise models as Gaussian (G), spatially-variant Gaussian (SV-G) and Poisson-Gaussian (P-G) distributions. We formulate the corresponding noises as follows,
\begin{equation}
	\begin{aligned}
		\mathbf{N}_\text{G}&\sim \mathcal{N}(\mathbf{0},\sigma^2_\text{r})\\
		\mathbf{N}_\text{SV-G}&\sim \mathbf{M}\odot\mathcal{N}(\mathbf{0},\mathbf{1})\\
		\mathbf{N}_\text{P-G}&\sim \mathcal{N}(\mathbf{0},\sigma^2_\text{s}\cdot \mathbf{L}+\sigma^2_\text{r})\\
	\end{aligned}
\end{equation}
obviously, Gaussian noise model is a specific case of P-G noise model when $\sigma_\text{s}=0$, and SV-G noise model is more fine-grain as applying a spatial map $\mathbf{M}$ to represent the variance of Gaussian in pixel-wise level, thus P-G noise model can be regarded as a specific case of SV-G by setting the map relative to image context $\mathbf{L}$. 

\begin{table*}[!t]
	\renewcommand\arraystretch{1.}
	\small
	\centering
	\caption{Quantitative comparisons of different real image denoising methods on several benchmarks. Particularly, the criteria of SIDD and DND benchmarks are obtained from the corresponding online server.}
	\begin{tabular}{l c | cc | cc | cc}
		\hline
		\hline
		\multirow{2}{*}{Method}&Params$\downarrow$&\multicolumn{2}{c|}{SIDD Validation}&\multicolumn{2}{c|}{SIDD Benchmark$^\dagger$}&\multicolumn{2}{c}{DND Benchmark$^\dagger$}\\
		&(M)&PSNR$\uparrow$&SSIM$\uparrow$&PSNR$\uparrow$&SSIM$\uparrow$&PSNR$\uparrow$&SSIM$\uparrow$\\
		\hline
		Noisy&-&23.66&0.4848&29.56&0.3347&29.84&0.7015\\
		RIDNet \citep{AnwarS2019ICCV}&1.5&38.77&0.9511&38.98&0.9076&39.24&0.9513\\
		VDN \citep{YueZ2019NeurIPS}&7.8&39.36&0.9562&39.49&0.9117&39.30&0.9493\\
		MIRNet \citep{ZamirS2020ECCV}&31.8&39.72&0.9586&39.80&0.9147&39.88&0.9543\\
		MPRNet \citep{ZamirS2021CVPR}&15.7&39.71&0.9586&39.80&0.9149&39.82&0.9540\\
		Uformer \citep{WangZ2022CVPR}&50.8&39.89&0.9594&39.97&0.9160&\underline{40.05}&\underline{0.9562}\\
		Restormer \citep{ZamirS2022CVPR}&26.1&\underline{40.02}&\underline{0.9603}&\underline{40.09}&\underline{0.9171}&40.03&\textbf{0.9564}\\
		NAFNet \citep{ChenL2022ECCV}&29.1&39.97&0.9599&40.04&0.9166&39.10&0.9495\\
		MSANet \citep{GouY2022NeurIPS}&8.6&39.56&0.9575&39.70&0.9131&39.65&0.9553\\
		CAT \citep{ZhengC2022NeurIPS}&25.8&40.01&0.9600&\underline{40.09}&0.9167&\underline{40.05}&0.9561\\
		MIRNetv2 \citep{ZamirS2022TPAMI}&5.9&39.84&0.9593&39.91&0.9154&39.86&0.9550\\
		ShuffleFormer \citep{XiaoJ2023ICML}&50.1&40.00&\underline{0.9603}&40.08&0.9168&40.01&0.9560\\
		GRL \citep{LiY2023CVPR}&19.8&39.89&0.9595&40.01&0.9161&39.76&0.9540\\
		ART \citep{ZhangJ2023ICLR}&25.7&39.96&0.9598&40.03&0.9164&\underline{40.05}&0.9557\\
		VIRNet \citep{YueZ2024TPAMI}&15.4&39.70&0.9586&39.78&0.9148&39.77&0.9533\\
		MambaIR \citep{GuoH2024ECCV}&23.2&39.89&0.9598&39.97&0.9164&39.83&0.9542\\
		Condformer (Ours)&27.0&\textbf{40.21}&\textbf{0.9612}&\textbf{40.23}&\textbf{0.9176}&\textbf{40.10}&\underline{0.9562}\\
		\hline
		\hline
	\end{tabular}
	\label{tab:SOTA_Real}
\end{table*}
In comparison of synthetic image denoising, several current state-of-the-art blind denoisers are selected, including: 

\noindent{\bf \em 1) Gaussian noise model drived}: DnCNN \citep{ZhangK2017TIP} and Restormer \citep{ZamirS2022CVPR}; 

\noindent{\bf \em 2) SV-G noise model drived}: VDN \citep{YueZ2019NeurIPS}, DRANet \citep{WuW2024PR} and VIRNet \citep{YueZ2024TPAMI}; 

\noindent{\bf \em 3) P-G noise model drived}: FBI-Denoiser \citep{ByunJ2021CVPR}, VBDNet \citep{LiangH2023PR}, CLIPDenoising \citep{ChengJ2024CVPR} and our Condformer. 

In addition, we re-train the Restormer \citep{ZamirS2022CVPR} and MambaIR \citep{GuoH2024ECCV} denoisers on P-G noise model with same settings as our Condformer for fair comparisons, marked as Restormer$^\dagger$ and MambaIR$^\dagger$. 

As reported in Table~\ref{tab:SOTA_Poisson-Gaussian}, quantitative comparisons on three public validation datasets with various noise levels show that, our Condformer achieves superior performances against other methods. On denoising accuracy, our Condformer obtains higher PSNR values especially when handling Poisson-Gaussian noises ($\sigma_\text{s}>0$) and shows slight disadvantages compared with the state-of-the-art Restormer trained on Gaussian noise model when handing Gaussian noises ($\sigma_\text{s}=0$). 	
On model generalization, under same experimental settings, our Condformer achieves excellent performances on all noise levels, instead Restormer$^\dagger$ and MambaIR$^\dagger$ show lower generalization on various noise levels. That is, our Condformer possesses a complete and conditional optimization space for divide-and-conquer, which is naturally stronger than any unconditional models. 

Furthermore, we also provide the qualitative visual comparisons of restoring the Poisson-Gaussian noisy images in Fig.~\ref{fig:visual results_Poisson-Gaussian}. Our Condformer can handle any degree of noisy corruptions, being capable of detail preservation and noise removal well. Especially, under severe corruptions, only our Condformer simultaneously recovers the details (such as textures of the helmet, edges of the windows), alleviates the distortions in smooth region and improves the fidelity of color. 

\begin{figure*}[!t]
	\centering
	\includegraphics[width=1\linewidth]{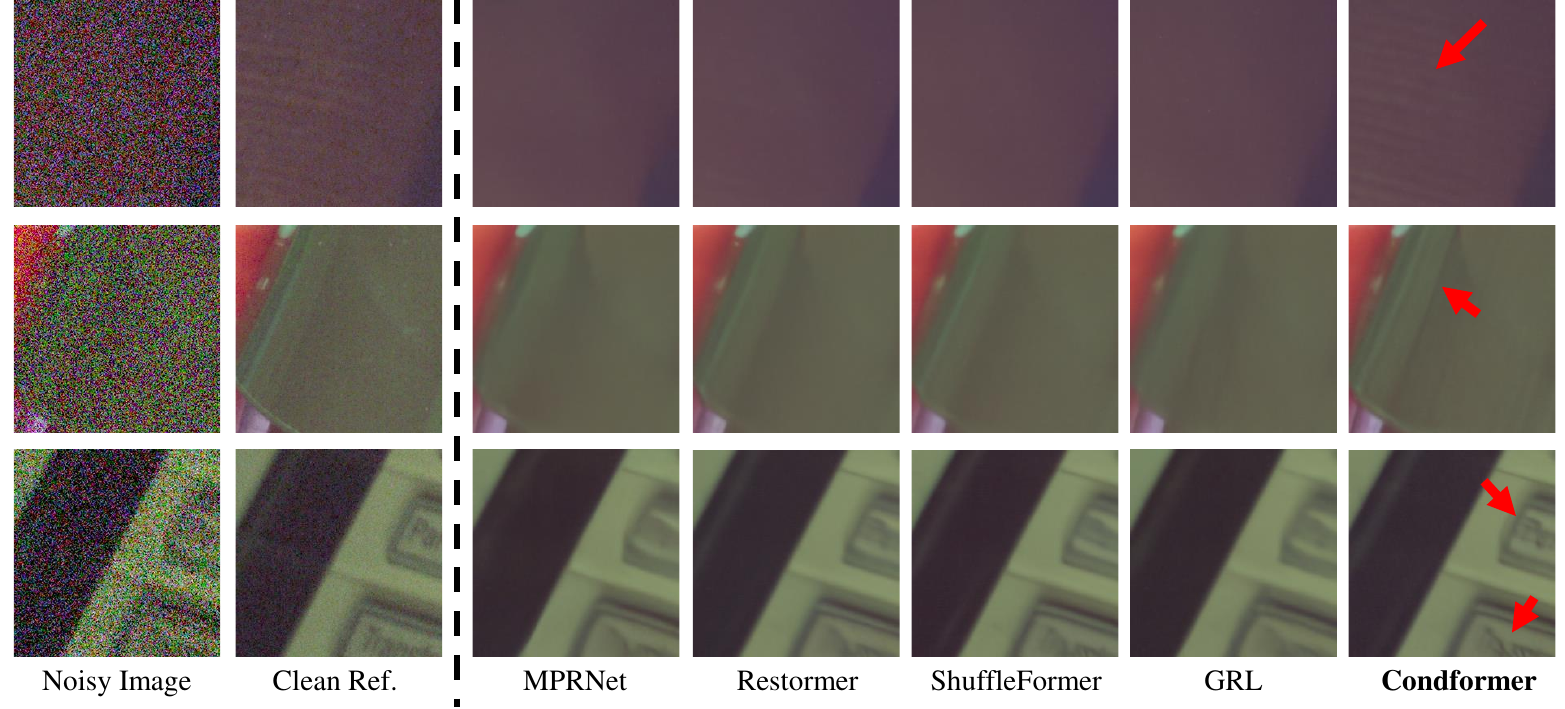}
	\caption{Qualitative comparisons on SIDD validation datasets. Our Condformer can preserve more details as accurately guide the model to remove the noises adaptively, instead other unconditional denoisers prone to handle the high-frequency details as undesired noises since the unknownability of noise level in training and testing phases.}
	\label{fig:SIDD-V_performance}
\end{figure*}

\begin{figure*}[!t]
	\centering
	\includegraphics[width=1\linewidth]{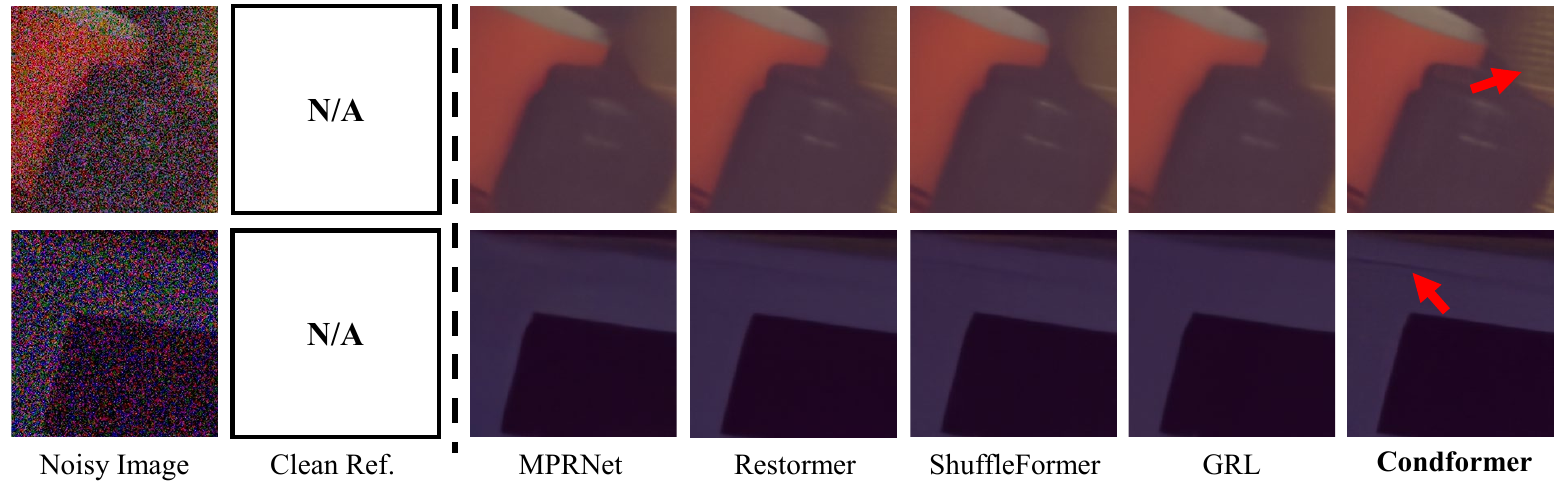}
	\caption{Qualitative comparisons on SIDD benchmark datasets. Especially under low-light imaging environments, noises with severe corruption prone to overwhelm the image contexts. In this case, only our Condformer could preserve more details, \eg, textures, edges and natural structures.}
	\label{fig:SIDD-B_performance}
\end{figure*}
\begin{figure*}[!t]
	\centering
	\includegraphics[width=1\textwidth]{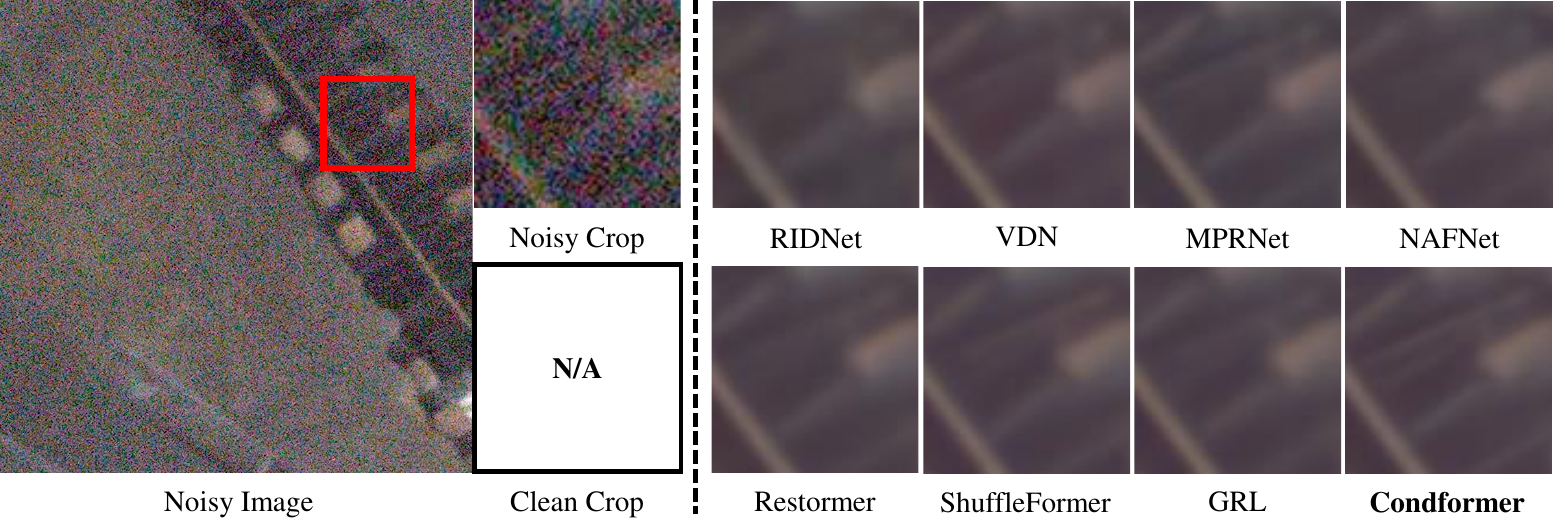}
	\caption{Qualitative comparisons on DND benchmark under normal-light imaging environment. Our Condformer preserves the edges and natural structures more distinctly than other denoisers.}
	\label{fig:DND_performance_1}
\end{figure*}
\begin{figure*}[!t]
\centering
\includegraphics[width=1\textwidth]{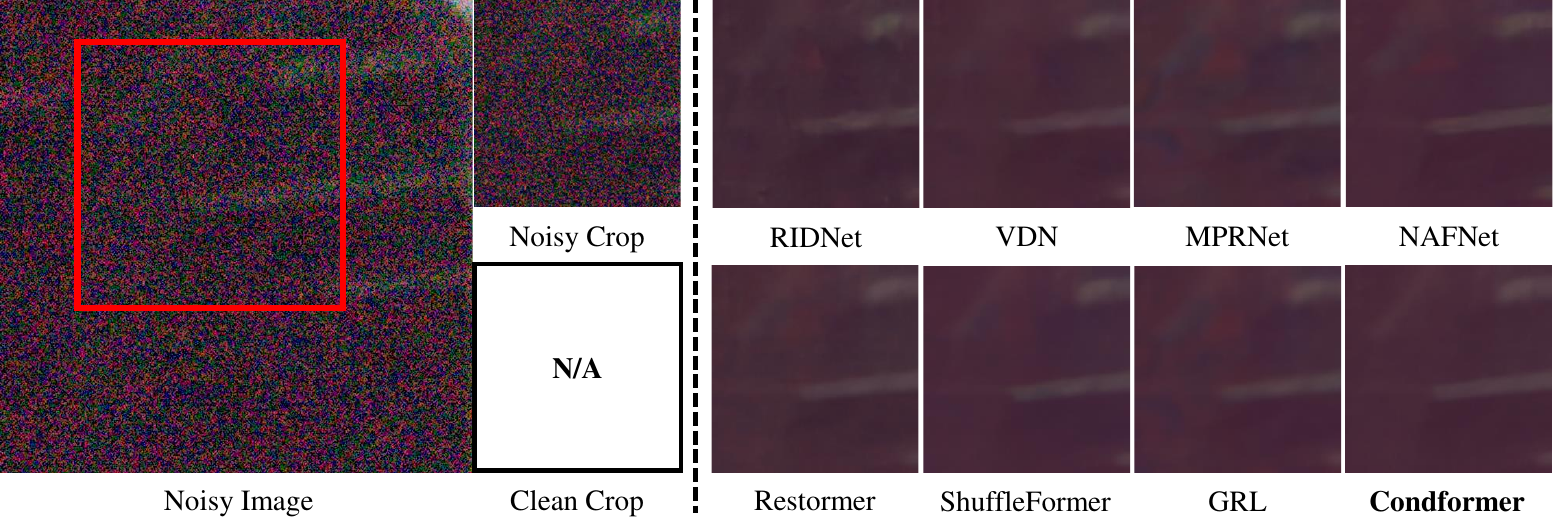}
\caption{Qualitative comparisons on DND benchmark under low-light imaging environment. Although noise interferes the image context restoration overwhelmingly under low-light environment, our Condformer remains fewer noises and can restore cleaner flatten regions than other denoisers.}
\label{fig:DND_performance_2}
\end{figure*}

\subsubsection{On real images}

In this section, we demonstrate the effectiveness of the proposed Condformer for real image denoising. We compare our Condformer with several state-of-the-art real image denoising methods on SIDD validation, SIDD benchmark, and DND benchmark datasets. We select several representative CNN-based denoisers and current state-of-the-art Transformer-based and Mamba-based denoisers, including: 

\noindent{\bf \em 1) CNN-based denoiser}: RIDNet \citep{AnwarS2019ICCV}, VDN \citep{YueZ2019NeurIPS},MIRNet \citep{ZamirS2020ECCV}, MPRNet \citep{ZamirS2021CVPR}, NAFNet \citep{ChenL2022ECCV}, MSANet \citep{GouY2022NeurIPS}, MIRNetv2 \citep{ZamirS2022TPAMI} and VIRNet \citep{YueZ2024TPAMI}; 

\noindent{\bf \em 2) Transformer-based denoiser}: Uformer \citep{WangZ2022CVPR}, Restormer \citep{ZamirS2022CVPR}, CAT \citep{ZhengC2022NeurIPS}, ShuffleFormer \citep{XiaoJ2023ICML}, GRL \citep{LiY2023CVPR} and ART \citep{ZhangJ2023ICLR}; 

\noindent{\bf \em 3) Mamba-based denoiser}: MambaIR \citep{GuoH2024ECCV}. 

As reported in Table~\ref{tab:SOTA_Real}, our Condformer achieves the highest PSNR and SSIM criteria over all the compared denoising methods. Note that, since the clean groundtruth images of SIDD and DND benchmarks are inaccessible, we upload the denoised results of all the considered methods on their online servers for testing.

Particularly, compared with other Transformer-based denoisers, our Condformer needs relatively lower complexity. For example, on SIDD validation dataset, our Condformer achieves 0.32dB gain of PSNR over Uformer with about half of its parameters. Even though Restormer is considered as the baseline of our Condformer without embedded noise prior in the self-attention module, our Condformer gets about 0.2dB gain of PSNR over it with similar computational complexities. In spite of training on a single SIDD-Medium dataset, our Condformer performs excellently on the out-of-distribution DND benchmark.

Qualitatively on the in-distribution SIDD validation and benchmark datasets, as shown in Fig.~\ref{fig:SIDD-V_performance} and Fig.~\ref{fig:SIDD-B_performance}, our Condformer can not only successfully remove the noises but also keep the details well, including textures, edges and natural structures. After conditional optimization with noise prior, our Condformer could handle each noisy image in a sensor-specific way, where the denoising intensity is proportional to the corruption degree of the noisy observation. Especially under low-light imaging environments, noises prone to overwhelm the image contexts as introducing higher ISO level and indirectly enlarging the shot noises. 

Furthermore, Fig.~\ref{fig:DND_performance_1} and Fig.~\ref{fig:DND_performance_2} show that our Condformer has excellent generalization ability as handling the out-of-distribution DND noisy images well, on both detail preservation and undesired noise removal. Especially under severe corruptions, our Condformer introduces lower distortions of color than other methods as shown in Fig.~\ref{fig:DND_performance_2}.

Note that, since the images have low visibility under low-light imaging environment, we employ image normalization on each denoised result for better visualization. 


\begin{table*}[!t]
	\renewcommand\arraystretch{1.}
	\small
	\centering
	\caption{Quantitative comparisons of different Transformer-based denoising methods on computational complexity. The FLOPs, Memory and Time are calculated when processing a single $512\times512$ sRGB image on a NVIDIA A800 GPU.}
	\begin{tabular}{l | cccc}
		\hline
		\hline
		\multirow{2}{*}{Method}&Parameter$\downarrow$&FLOPs$\downarrow$&Memory$\downarrow$&Time$\downarrow$\\
		&(M)&(G)&(G)&(s)\\
		\hline
		Uformer \citep{WangZ2022CVPR}&50.8&343.1&2.89&0.21\\
		Restormer \citep{ZamirS2022CVPR}&26.1&564.0&3.81&0.36\\
		CAT \citep{ZhengC2022NeurIPS}&25.8&543.5&10.83&15.76\\
		ShuffleFormer \citep{XiaoJ2023ICML}&50.5&344.4&2.97&0.23\\
		GRL \citep{LiY2023CVPR}&19.8&5012.3&12.38&10.45\\
		ART \citep{ZhangJ2023ICLR}&25.7&542.8&3.66&0.41\\
		Condformer (Ours)&27.0&565.2&3.81&0.37\\
		\hline
		\hline
	\end{tabular}
	\label{tab:SOTA_complexity}
\end{table*}

\begin{table*}
	\centering
	\small
	\renewcommand\arraystretch{1.}
	\caption{Ablation study of Condformer with different noise priors on SIDD Validation dataset.}
	\begin{tabular}{c c c c | cccc}
		\hline
		\hline
		\multirow{2}{*}{\tabincell{c}{CondSA}}&\multicolumn{2}{c}{Noise Prior}&\multirow{2}{*}{LFM}&Parameter$\downarrow$&FLOPs$\downarrow$&Time$\downarrow$&PSNR$\uparrow$\\
		&Value&Location&&(M)&(G)&(s)&(dB)\\
		\hline
		\ding{55}&-&-&-&26.11&563.96&0.361&39.97\\
		\hdashline
		\ding{51}&$(0, 0)$&($\mathbf{Q}, \mathbf{K}$)&\ding{55}&26.13&564.04&0.363&39.96\\
		\ding{51}&$(\hat\sigma_\text{s}, \hat\sigma_\text{r})$&($\mathbf{Q}, \mathbf{K}$)&\ding{55}&26.74&564.22&0.368&40.09\\
		\ding{51}&$(\hat\sigma_\text{s}, \hat\sigma_\text{r})$&($\mathbf{Q}, \mathbf{K}$)&\ding{51}&27.02&565.35&0.368&40.21\\
		\ding{51}&$(\sigma_\text{s}, \sigma_\text{r})$&($\mathbf{Q}, \mathbf{K}$)&\ding{51}&26.41&565.17&1.012&40.23\\
		\ding{51}&$(\sigma_\text{s}, \sigma_\text{r})$&($\mathbf{Q}, \mathbf{K}, \mathbf{V}$)&\ding{51}&27.16&565.95&1.014&40.19\\
		\hline
		\hline
	\end{tabular}
	\label{tab:ablation_Condformer}
\end{table*}
\subsubsection{On computational complexity}
In this section, we mainly conduct a comparison of our Condformer and other Transformer-based denoising methods on several computational complexity criteria. In detail, the number of parameters represents the model size for transmission and storage necessaries, FLOPs and Time indicate the time complexity of model, and Memory is the memory usage when running model on a GPU device, indicating the threshold level of training and inference resources.

The last three criteria are calculated when processing a single $512\times 512\times 3$ sRGB image input on a NVIDIA A800 GPU. To avoid randomness, the running time is averaged on handling 100 images with \texttt{torch.cuda.Event} timer. As reported in Table~\ref{tab:SOTA_complexity}, our Condformer shows relatively trade-off on the complexities, which is a resource-friendly method.



\subsection{Model Analysis}
In this section, we mainly investigate the effects of our Condformer and LoNPE modules with experimental analysis.

\begin{table*}[!t]
	\centering
	\small
	\renewcommand\arraystretch{1.}
	\caption{Investigation on LoNPE algorithm with different patch-sampling hyperparameters. Particularly, $\mathcal{O}$, $m/n$ and $\lambda_\mathcal{S}$ denote the sampling size, ratio and index, respectively. Real and synthetic scenes are considered on the public SIDD-Medium and Urban100 datasets, respectively.}
	\begin{tabular}{c c c| c c | c c c}
		\hline
		\hline
		\multicolumn{3}{c|}{LoNPE Algorithm}&\multicolumn{2}{c|}{SIDD-Medium}&\multicolumn{3}{c}{Urban100}\\
		$\mathcal{O}$&$m/n$&$\lambda_\mathcal{S}$&CV$\downarrow$&Time$\downarrow$&CV$\downarrow$&RMSE$\downarrow$&Time$\downarrow$\\
		\hline
		$8\times8$&$5$\%&\ding{55}&0.213&0.39&0.533&0.125&0.07\\
		\hdashline
		$8\times8$&$5$\%&\ding{51}&0.048&0.41&0.248&0.021&0.09\\
		$8\times8$&$10$\%&\ding{51}&0.040&0.52&0.233&0.020&0.11\\
		\hdashline
		$16\times16$&$10$\%&\ding{51}&0.031&0.81&0.241&0.020&0.17\\
		$16\times16$&$20$\%&\ding{51}&0.036&1.40&0.325&0.031&0.28\\
		\hdashline
		$32\times32$&$10$\%&\ding{51}&0.041&1.12&0.414&0.045&0.24\\
		\hline
		\multicolumn{3}{c|}{LoNPE Network}&0.032&0.01&0.280&0.025&0.01\\
		\hline
		\hline
	\end{tabular}
	\label{tab:ablation_LoNPE}
\end{table*}

\subsubsection{Investigation on Condformer}

Aiming at guiding the model to learn from image prior and noise prior separately, the core of our Condformer is embedding the noise prior effectively in the latent space for conditional optimization. Therefore, we conduct the ablation study in Table~\ref{tab:ablation_Condformer} using the same training settings as those used for the full model.

Since the core of our Condformer is constructing CondSA blocks in the latent space, we first train a baseline model which is same sa Restormer and achieves PSNR of 39.97dB on SIDD validation dataset. Subsequently, to ensure that the introduction of an auxiliary vector does not cause any performance deviation, we then build a void CondSA module by embedding a zero vector, and observe no improvement over the baseline model. However, after embedding the estimated noise prior $(\hat \sigma_\text{s},\hat \sigma_\text{r})$ from LoNPE network through a naive concatenation with the CondSA input tensor, the model achieves 0.13 PSNR gain, demonstrating the positive effect of introducing the noise prior. Furthermore, to learn an effective representation of the correlation between-in the intermediate image features and noise prior, a LFM module is designed for correlation representation of noise prior and image features. The result shows that an effective fusion module can significantly boost the contribution of the noise prior. As reported in Table~\ref{tab:ablation_Condformer}, the 4th model achieves higher performances than the afore three models, especially gains more than 0.24dB of PSNR against the baseline. 

Since noise prior is introduced as an embedding vector in CondSA, we prefer to embed it only into the query and key tensors and use the calculated attention map to enhance the value tensor. Compared to embedding the noise prior into all tensors, this design achieves higher denoising performance with lower computational complexity.

Moreover, although the precision of noise prior estimation might affects the denoising results, the Condformer with estimated noise prior $(\hat \sigma_\text{s},\hat \sigma_\text{r})$ is slightly inferior to the upper bound (the one with groudtruth $(\sigma_\text{s},\sigma_\text{r})$ obtained via LoNPE algorithm), but costs fewer time complexities. That is because the noise prior is a low-dimensional vector as easier to learn via a simple network, instead the LoNPE algorithm needs to calculate the numerical solutions iteratively which is time-consuming.

\subsubsection{Investigation on LoNPE}
As described in Section~\ref{sec:LoNPE}, our LoNPE algorithm is based on the statistical analysis of the sampled local smooth patches from a single noisy raw image. From Eq.(\ref{eq:LoNPE}), the noise prior is estimated on $m$ local patches with statistical mean $\{L_i\}^m_{i=1}$ and variance $\{\sigma^2_i\}^m_{i=1}$. Particularly, from Eqs.(\ref{eq:LoNPE_mean})-(\ref{eq:LoNPE_var}), the precision of statistical values $L_i$ and $\sigma^2_i$ are highly relied on the spatial size of samples $\mathcal{O}$, and the precision of parameter estimation is highly relied on the number of samples $m$ to ensure a large $\textit{Rank}([\mathbf{L}, \mathbf{1}])$. 

We investigate the effects of sampling size $\mathcal{O}$ and the sampling ratio $m/n$ in Table~\ref{tab:ablation_LoNPE}, focusing on the stability of noise prior estimation by analyzing the Coefficient of Variation (CV) in scenarios where groundtruth for the noise prior is unavailable. Our analysis is conducted on the SIDD-Medium raw-domain training dataset. Specifically, since each instance in SIDD-Medium contains only two frames, we augment them using flips and rotations to compute the CV criterion. Additionally, we synthesize noisy samples by randomly generating noise from a Poisson-Gaussian distribution with $(\sigma_\text{s},\sigma_\text{r})$ set to (0.05, 0.02), (0.1, 0.05) and (0.2, 0.1). Root Mean Square Error (RMSE) criterion is calculated to evaluate the accuracy of noise prior estimation. Beside, the running time is reported in Table~\ref{tab:ablation_LoNPE} to assess the efficiency of various sampling settings. 

Our observations reveal that increasing the patch size $\mathcal{O}$ and sampling ratio $m/n$ would negatively impact the noise prior estimation as introducing more image priors, resulting in reduced stability (higher CV) and accuracy (higher RMSE). larger patches yield more precise statistical values but are more likely to include image priors, such as edges and textures, which interfere with the estimation of the noise prior. Moreover, we observe that the CV values on the SIDD-Medium dataset are significantly lower than those on Urban100, due to the lower diversity of images in SIDD-Medium. This further highlights the detrimental effect of image priors on noise prior estimation.
Empirically, we select $\mathcal{O}=16\times16$ and $m/n=10$\%, as this configuration achieves a relatively better balance for estimating the noise prior across both real and synthetic scenes.

\begin{figure}
	\centering
	\includegraphics[width=1\linewidth]{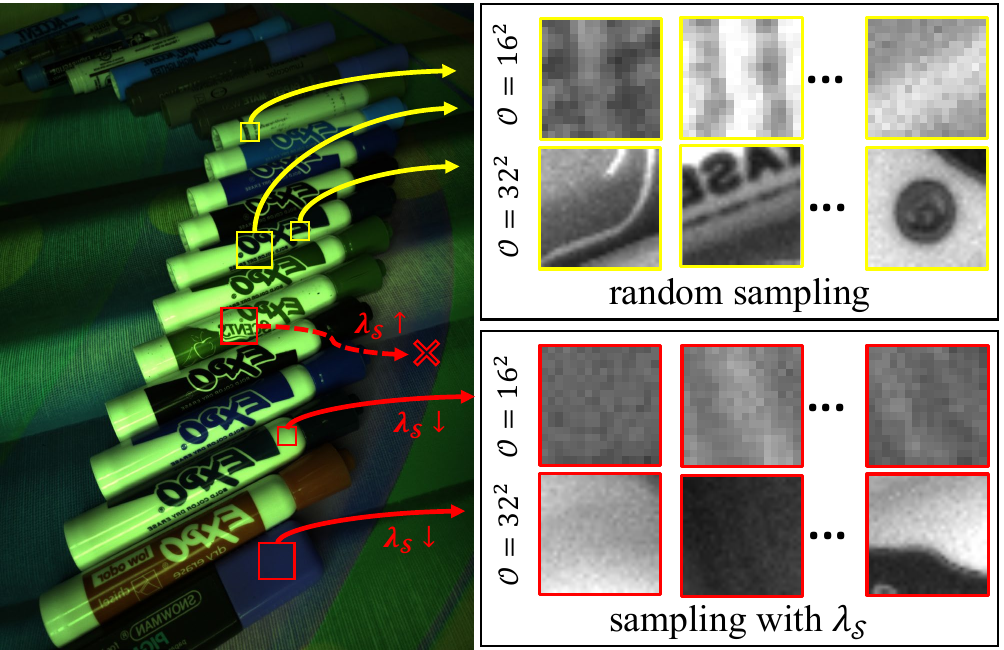}
	\caption{Visualization of local patch sampling with $\lambda_\mathcal{S}$. The local smoothness criterion $\lambda_\mathcal{S}$ can effectively filter the smooth patches to eliminate the interference of image prior to noise prior estimation, instead random sampling might introduce high-frequency patches and cause inaccuracy of local luminance calculation.}
	\label{fig:patch sampling LoNPE}
\end{figure}
Besides, to further eliminate the image prior in each patch, we introduce a local smoothness criterion $\lambda_\mathcal{S}$ to filter the smooth patches from the original patch pools $\{I_i\}^n_{i=1}$, which can effectively boost the accuracy of noise prior estimation as reported in Table~\ref{tab:ablation_LoNPE}. To further demonstrate the effectiveness of local smoothness criterion, we record the sampled local patches in Fig.~\ref{fig:patch sampling LoNPE}. It is obvious that random sampling local patches from the raw image inevitably captures the numerous high-frequency details, which would interfere the estimation of noise prior because of the nonnegligible deviation on calculating the local luminance in Eq.(\ref{eq:LoNPE_mean}). Instead, by employing the local smoothness criterion $\lambda_\mathcal{S}$, it is easy to filter the smooth patches from the original patch pools, which plays significant role on eliminating the image prior and help estimating noise prior.

\section{Conclusion and Discussions}\label{sec:conclusion}
In this paper, by rethinking the real image denoising task and revisiting the formation model of raw camera sensor noises, we have generalized a principle of the independence of image prior and noise prior. This principle guides an alternative conditional optimization to tackle the limitations of existing learning-based unconditional denoising methods. At the algorithmic level, we have presented a novel Condformer architecture, which effectively embeds the noise prior into the self-attention module. The noise prior is explicitly estimated using our LoNPE algorithm or network. Extensive experiments confirm the advantages of conditional optimization with noise prior, demonstrating that the proposed LoNPE and Condformer achieve superior performance on both synthetic and real noise statistics and image denoising tasks, respectively. 

Nonetheless, there are other factors affecting imaging conditions that are relevant for noise statistics analysis, including aperture, sensor size/type and \etc. Besides, due to the defective sensor and circuits technology, the formation model of raw sensor noise is actually more sophisticated than the Poisson-Gaussian noise model. For instance, read noise might follows a heavy-tailed Cauchy distribution \citep{WeiK2022TPAMI} in extremely low-light environments, dark shading \citep{FengH2024TPAMI} can result from sensor non-uniformity, and more sophisticated noise models are emerging \citep{CaoY2023CVPR}. Therefore, future work should explore the estimation of noise prior under different imaging factors and in more complex scenarios, aiming to further enhance noise estimation and denoising performance.

\section*{Acknowledgments}
This work was supported in part by the National Natural Science Foundation of China under Grant 62202056, and the Fundamental Research Funds for the Central Universities under Grant 2243100002.

\section*{Data Availability}
As introduced in Section~\ref{sec: datasets}, the datasets that support the findings of this study are openly available in the cited references. 
Our repository is provided in \url{https://github.com/YuanfeiHuang/Condformer}.

\bibliography{submission}

\end{document}